\documentclass[sigconf,authorversion,screen,nonacm]{acmart}

\usepackage{booktabs}
\usepackage{url}
\usepackage{graphicx}
\usepackage{xcolor}
\usepackage{amssymb}
\usepackage{amsmath}
\usepackage{paralist}
\usepackage{multirow}
\usepackage{subfig}

\newcommand{\specialcell}[2][c]{
	\begin{tabular}[#1]{@{}l@{}}#2\end{tabular}}

\setcopyright{none}

\acmConference[ASIACCS'19]{ACM ASIACCS Conference}{July 7-12, 2019}{Auckland, New Zealand} 
\acmYear{2019}
\copyrightyear{2019}

\acmArticle{4}
\acmPrice{15.00}

\begin{document}
\title[HADES-IoT: A Practical Host-Based Anomaly Detection System for IoT Devices]{HADES-IoT: A Practical Host-Based Anomaly Detection System for IoT Devices (Extended Version)}

\author{Dominik Breitenbacher}
\affiliation{
    \institution{Singapore University of Technology and Design}
}
\email{dbreitenbacher@gmail.com}
\author{Ivan Homoliak}
\affiliation{
    \institution{Singapore University of Technology and Design}
}
\email{ivan\_homoliak@sutd.edu.sg}
\author{Yan Lin Aung}
\affiliation{
    \institution{Singapore University of Technology and Design}
}
\email{linaung\_yan@sutd.edu.sg}
\author{Nils Ole Tippenhauer}
\orcid{0000-0001-8424-2602}
\affiliation{
    \institution{CISPA Helmholtz Center for Information Security}
}
\email{tippenhauer@cispa.saarland}
\author{Yuval Elovici}
\affiliation{
    \institution{Singapore University of Technology and Design}
}
\email{yuval\_elovici@sutd.edu.sg}

\renewcommand{\shortauthors}{}

\begin{abstract}
Internet of Things (IoT) devices have become ubiquitous and are spread across many application domains including the industry, transportation, healthcare, and households.
However, the proliferation of the IoT devices has raised the concerns about their security, especially when observing that many manufacturers focus only on the core functionality of their products due to short time to market and low cost pressures, while neglecting security aspects.
Moreover, it does not exist any established or standardized method for measuring and ensuring the security of IoT devices.
Consequently, vulnerabilities are left untreated, allowing attackers to exploit IoT devices for various purposes, such as compromising privacy, recruiting devices into a botnet, or misusing devices to perform cryptocurrency mining.

In this paper, we present a practical Host-based Anomaly DEtection System for IoT (HADES-IoT) that represents the last line of defense.
HADES-IoT has proactive detection capabilities, provides tamper-proof resistance, and it can be deployed on a wide range of Linux-based IoT devices.
The main advantage of HADES-IoT is its low performance overhead, which makes it suitable for the IoT domain, where state-of-the-art approaches cannot be applied due to their high-performance demands.
We deployed HADES-IoT on seven IoT devices to evaluate its effectiveness and performance overhead.
Our experiments show that HADES-IoT achieved 100\% effectiveness in the detection of current IoT malware such as VPNFilter and IoTReaper; while on average, requiring only $5.5\%$ of available memory and causing only a low CPU load.
\end{abstract}


\maketitle

\section{Introduction}

In recent years, the number of IoT devices connected to the Internet reached seven billion~\cite{lueth2018iotanalytics} and is expected to grow.
Gartner estimates that more than 20 billion IoT devices will be connected to the Internet by 2020~\cite{meulen2017gartner}.
There are two reasons for this trend: ubiquitous Internet connectivity and decreasing cost of embedded computing technology.
Currently, IoT devices are utilized in various application domains including healthcare, transportation, entertainment, industrial control, smart buildings/homes, and others. 
Nevertheless, the advent of the IoT has brought challenges in many areas, including data storage, maintenance, and particularly, privacy and security ~\cite{Serpanos:2013:SCE:2435227.2435262}, \cite{hulme2012}, \cite{newell2016}.
Many IoT devices are designed with a particular purpose in mind, so their software and hardware is solely chosen to satisfy the requirements of core functionalities, e.g., using just the fastest processor needed to meet certain real-time constraints and nothing faster.
When it comes to security, the preference is put on fast time to market and budget constraints at the expense of more expensive and possibly more comprehensive security solutions. 
Therefore, embedded technology results in a trade-off between cost and security~\cite{Serpanos:2013:SCE:2435227.2435262}; real-time requirements, computing capabilities, and energy consumption are also part of this trade-off.

For these reasons, IoT devices are often released with serious vulnerabilities, and this issue is further exacerbated by the fact that IoT devices are, in many cases, exposed on the Internet, and thus easily accessible to attackers.
Once a vulnerable device is compromised, it can be exploited for various purposes.
Currently, the most common scenario is that the compromised device becomes a part of a botnet that performs DDoS attacks~\cite{cimpanu2016renting}, \cite{gooding2018renting}, \cite{nuvias2017renting}.
However, recently, attackers have started to utilize IoT devices for cryptocurrency mining as well~\cite{canellis2018cryptocurr}, \cite{merces2018cryptocurr}.
Most IoT devices were exploited due to operations security (OPSEC) issues, such as the use of weak or default passwords~\cite{antonakakis2017understanding}, \cite{owasp2018iot}; however, they are also exploited due to buffer overflow, command injection, etc., \cite{trendmicro2018vpnfilter}.
These practical examples raise the question of how to achieve greater security in IoT devices while minimizing the requirements on cost and the utilization of computational resources?
One way to raise the bar against attackers is to improve preventive OPSEC countermeasures, such as employing strong passwords and conservative access control.
However, in the case of more sophisticated attacks that cannot be prevented by OPSEC countermeasures, IoT devices must be protected by other dedicated means, such as host-based intrusion detection systems or IoT-specific antivirus systems.
As discussed above, there might not be enough economic incentives for companies developing low-cost competitive IoT devices to invest in security countermeasures that are also computationally expensive.
On the other hand, as we have learned after many years of discussion regarding the security of IT and software in general, serious vulnerabilities can remain hidden due to the complexity of modern systems and the inherent difficulty to reveal such vulnerabilities.
For example, even a securely written application might be vulnerable due to bugs in third-party libraries, compiler, or even an operating system.

Given the abovementioned constraints, we make the following fundamental observation: in contrast to general computing devices such as laptops or mobile devices, IoT devices have, by design, a well-defined and stable functionality.
Moreover, this functionality is usually provided by a small set of system processes that have mostly stable behavior, as IoT devices are rarely updated. 
However, when an IoT device is compromised by malware, its behavior changes significantly.
Therefore, intrusion detection and anomaly detection systems are promising options for securing IoT devices.
Although, these systems might be deployed outside of the device and perform inspection of network traffic~\cite{homoliak2016intrusion}, they might be evaded by various payload-based~\cite{fogla2006polymorphic,vigna2004testing} and non-payload-based obfuscations~\cite{boltz2010new,ih2019nonpayload}. 
Hence, we argue that behavioral changes are best observed from ``within'' a device, for instance, by monitoring which processes are running and what actions they perform. 
For this reason, we consider host-based behavior analysis as an effective last line of defense.
In particular, we aim at process-based adaptation of an anomaly detection approach that profiles the normal behavior of an IoT device and strictly detects all anomalies deviating from this profile.

In this paper, we propose a lightweight Host-based Anomaly DEtection System for IoT devices (HADES-IoT) that monitors process spawning and stops any unauthorized program before its execution, thus providing proactive detection and prevention functionalities.
To achieve real-time detection, HADES-IoT has been developed in the form of a loadable kernel module of Linux-based operating systems.
Such a design decision allows us to make HADES-IoT tamper-proof resistant against an attacker (with superuser privileges) trying to disable it.
Since IoT devices are significantly resource constrained, HADES-IoT was created in a lightweight fashion, ensuring that primary functionality provided by the device is not affected.
Also, as most IoT devices are based on the Linux operating system~\cite{cho2017business}, HADES-IoT supports various types of IoT devices.

\vspace{0.1cm}
\noindent \textbf{Contributions:} In summary, our contributions are as follows:
\begin{compactitem}
    
	\item We present a novel host-based anomaly detection and prevention approach that is based on whitelisting legitimate processes on an IoT device. 	
	    
	\item We develop a proof-of-concept of our approach and evaluate its effectiveness on several IoT devices, including very resource-restricted devices.
    
	\item We show that HADES-IoT can be easily adapted to any Linux kernel version, which makes it generic.
    
	\item We demonstrate that HADES-IoT is resilient against attacks that focus on disabling its protection mechanisms, and thus providing tamper-proof feature.
	
\end{compactitem}
\vspace{0.1cm}

The rest of this paper is organized as follows.
We describe the problem statement in Section~\ref{sec:problem_statement}.
In Section~\ref{sec:preliminaries} we explain preliminaries of our work.
Then in Section~\ref{sec:hades_approach}, we explain details of our approach; we perform evaluation in Section~\ref{sec:evaluation}. 
Section~\ref{sec:discussion} discusses limitations and possible extensions of the approach. Section~\ref{sec:related_work} is dedicated to related work and Section~\ref{sec:conclusion} concludes the paper.

\section{Problem Statement}
\label{sec:problem_statement}

The main objective of this work is to propose a security solution that protects the bulk of the existing IoT devices against remote exploitation of any vulnerabilities (including zero-day ones).

\subsection{Assumptions}
\label{sec:prob_stat_assumptions}

This work is aimed at Linux-based IoT devices.
We argue that according to~\cite{cho2017business}, the market share of Linux-based IoT devices is over 80\%, and hence in this paper we are targeting the vast majority of existing IoT devices.
Furthermore, we assume that (1) all of the executables installed on an IoT device are benign, and (2) an attacker does not tamper with the device either before or during the bootstrapping of our proposed approach.
However, we assume that the default executables of an IoT device may contain a vulnerability enabling execution of arbitrary executable binaries -- either binaries that already exist on a device or binaries delivered by an attacker who exploits such a vulnerability.
Finally, to the best of our knowledge, it is not common practice for manufacturers to enable security features, such as SELinux, Auditd, or Access Control Lists on their IoT devices mainly due to performance reasons. 
Therefore, we assume that our approach is the only security solution deployed on an IoT device.

\subsection{Attacker Model}
\label{sec:prob_stat_attacker_model}

We assume that an IoT device is protected by our approach and is connected to the Internet (i.e., has a public IP address). 
Therefore, an attacker is able to find it using a custom scanner or any publicly available services such as Shodan.\footnote{\url{https://www.shodan.io/}}
The attacker is also capable to scan an IoT device in order to reveal open ports and identify running network services.
Finally, we consider that an attacker is capable of exploiting a (potential zero-day) vulnerability on an IoT device through one of the network services running on the device.
This can be accomplished, for example, by brute-forcing passwords or using default passwords to services such as Telnet or SSH.
However, according to the creator of Sora and Owari IoT malware~\cite{anubhav2018newskysecurity}, Telnet service is currently abused by attacks brute-forcing passwords to such a large extent that there is a growing trend to target services that are not as heavily exploited.
Therefore, the developers of IoT malware have begun to integrate exploit scanning tool into their malware and scanners more frequently.
Such vulnerability exploitation provides them with an additional attack vector (beyond Telnet misuse), and improves their ability to compromise exposed and vulnerable IoT devices, even when the devices are protected with strong passwords.
For example, VPNFilter takes advantage of 19 vulnerabilities, enabling this malware to compromise about 70 models from 10 different vendors~\cite{trendmicro2018vpnfilter}.

To accurately reflect real-world attack scenarios, we further consider that once an attacker ``is inside'' an IoT device, the attacker is granted superuser privileges, since it is quite common for an IoT device to have just a superuser account.

\subsection{Requirements for an IoT Defense Solution}
\label{sec:prob_stat_requirements}

In this section, we specify desirable properties and requirements for a host-based IoT defense solution resistant against the abovementioned attacker models.
In particular, a defense solution should meet the following requirements:
\begin{compactitem}
    \item[\textbf{Real-Time Detection:}] Since we aim to prevent any unknown action on an IoT device, a defense solution must be capable of detecting any unknown program upon its execution.
        
    \item[\textbf{Lightweight Overhead:}] IoT devices are extremely resource constrained, and thus provide only limited processing and storage resources. 
    Therefore, it is not possible to utilize conventional security approaches used in PC environments (e.g.,~machine learning or complex heuristics approaches).
    With this in mind, a defense solution should be conservative in terms of resource consumption and should only utilize existing dependencies.
    
    \item[\textbf{Tamper-Proof Protection:}] 
    Since the attacker has superuser privileges, he may terminate or bypass a defense solution deployed on an IoT device.
    Therefore, a defense solution should be resilient against such a powerful attacker.         
        
    \item[\textbf{Wide Coverage:}] 
	It is important to protect a wide range of IoT devices (e.g.,~printers, IP cameras, Wi-Fi routers, etc.), taking into account the fact that a significant portion of the existing IoT devices is already considered legacy and moreover may lack updates from manufacturers.
    
    \item[\textbf{Independence:}] 
    The deployment of a defense solution must not be dependent on a manufacturer; both the user and a manufacturer must be capable of deploying the defense solution.
	
    \item[\textbf{Ease of Bootstrapping:}] 
    With regard to the deployment of a defense solution mentioned above, we further argue that a defense solution should be capable of being deployed with minimal effort and should not require recompilation of the kernel of the IoT device's OS.    

\end{compactitem}

\subsection{Design Problems and Options}
\label{sec:prob_stat_design_space}

Based on the defined requirements, we analyzed our options for developing a defense solution and identified additional constraining factors that should also be considered as well.
Initially, we conjectured that the most straightforward option is to utilize features provided by Linux, such as \emph{KProbe}\footnote{\url{https://www.kernel.org/doc/Documentation/kprobes.txt}} or \emph{inotify}.\footnote{\url{http://man7.org/linux/man-pages/man7/inotify.7.html}}
This would provide us with the support of the Linux kernel and we would be able to base the defense solution on information provided by these features.
However, we examined several IoT devices (see Table~\ref{tab:tab_devices}) and found that these features are not supported in any of them. 
Another design aspect we considered is the lightweight complexity and low resource requirements of the defense solution.
In particular, the most important resources for a defense system are CPU and memory.
We  measured the normal utilization of these resources by the IoT devices in this study and found out that while there is a reasonable reserve of CPU utilization, the CPU performance is often low.
Therefore, we must ensure that the defense solution minimizes CPU consumption only to the extent needed.
Otherwise, other applications could be affected (e.g., delayed in performing their normal actions), which could further deteriorate the performance and availability of the device.
We also observed that there is only a small amount of free memory on IoT devices (e.g., in some cases lower than 2MB), however not all of the free space can be utilized for a defense solution, since other applications might rely on it. 

Since the challenge is to detect unknown processes in real-time upon their spawning, the Linux process scheduler is another limiting factor.
In the user space environment, processes compete for the CPU, and the process scheduler makes decisions regarding the CPU and time allocations for the processes.
Therefore, if a defense solution were implemented in user space, there is no guarantee that it would be running when a new process is spawned; hence, malicious processes might be missed or detected too late.
This issue could be na\"{i}vely mitigated by setting the highest priority to the defense solution, so the scheduler would prefer it over others;
however, this does not resolve the above issue, since the attacker possesses the same capabilities as a defense solution, and thus malicious processes would compete with the defense solution, having an equal chance of being selected by the scheduler.
This issue is exacerbated by the fact that IoT devices are often equipped with just one CPU core that contains a single processing thread.
Therefore, to ensure that none of the newly spawned processes is missed, the defense solution cannot be dependent on the scheduler and its planning algorithm, and must be based on a technique that is always triggered upon spawning a new process.

System call interception is a suitable technique capable of addressing the issue of execution priority.
System call interception inserts a code with the defender's desired functionality between the caller's invocation of a system call and the system call itself. Thus, with an appropriate set of intercepted system calls, this technique enables all new processes to be ``caught'' upon their spawning and checked to determine whether they are authorized to run.
In general, there are two options for performing system call interception.
The first option is libC library hooking, which is performed in the user space, and the second option is the interception of system calls through a loadable kernel module (LKM) running in kernel space.
Although libC hooking is easier to develop with more freedom compared to creating a loadable kernel module, it is not suitable for our case, because the IoT environment is diverse, and each manufacturer uses custom Linux that can be compiled with various (or custom) libraries and their versions (including libC).
Attackers address these issues by compiling their malware statically\footnote{Meaning that all required libraries are included within the malware's binary.} in order to cover as many IoT devices as possible. 
This fact renders libC hooking unusable for the detection of the majority of malware.
Because of this and the fact that it can fulfill most of the requirements defined in Section~\ref{sec:prob_stat_requirements}, we identify LKM as the most feasible solution.

\section{Preliminaries}
\label{sec:preliminaries}

Based on the analysis of possible options to achieve our goals, we decided to follow the LKM option that utilizes the system call interception technique.
In this section, we explain preliminaries related to these techniques and provide a few examples.

\subsection{System Calls}
On Linux systems, every process starts in a non-privileged mode~\cite{bellevue2006system}.
In this mode, a process is restricted, and only capable of using the memory space assigned to it by the Linux system. 
Access to the memory space of other programs or kernel is thus prohibited and results in raising an exception.
This ensures that only the Linux kernel has control of all of the resources.
Therefore, when a program wants to access a resource outside of its allocated memory space, such as reading a file or executing another process, it must perform a call to the kernel, which is realized by system calls.
Every system call has a unique numeric ID and represents a request for a specific operation provided by the kernel.
Currently, a Linux kernel provides the user with more than 300 different system calls~\cite{kerrisk2017linux}.
Upon a system call invocation, a software interrupt is raised, which results in switching into the system mode.
In the system mode, the process is granted root privileges.
This enables the Linux system to perform restricted actions.
Next, the requested operation is performed by the kernel of the Linux system, which ensures that the user space process does not interfere with the restricted resources.

\subsubsection*{\textbf{System Calls that Spawn New Processes}}
\label{sec:prelim_sycalls_spawn_new_proc}

Linux kernel provides three system calls that spawn new processes: 
\begin{compactitem}
	\item \textbf{Fork().} Upon its invocation, \emph{fork} creates a new process (i.e.,~child) by duplicating the calling process (i.e.,~parent). 
	Immediately after a call of \emph{fork}, parent and child run in separate memory spaces, but the spaces contain the same content.
	\item \textbf{Vfork().} Similarly to \emph{fork}, \emph{vfork} creates a child process of the calling process, but in contrast to \emph{fork}, the child and parent share the memory space after invocation. 
	Moreover, the parent is suspended until either the child terminates normally or it calls the \emph{execve} system call.
	As a result of this behavior, \emph{vfork} is often used in performance-sensitive programs in which a child immediately makes a call to \emph{execve} after \emph{vfork} has been invoked.
	\item \textbf{Clone().}  In contrast to the previous system calls, \emph{clone} can create a new thread, in addition to a new process.
	The \emph{clone} is more versatile compared to \emph{fork}, thus libC library implementations like \emph{glibc} provide a \emph{fork} wrapper which internally calls \emph{clone}.

\end{compactitem}

\begin{figure}\centering 	
	\centering
	
	\vspace{-0.5cm}    
	\includegraphics[width=0.85\linewidth]{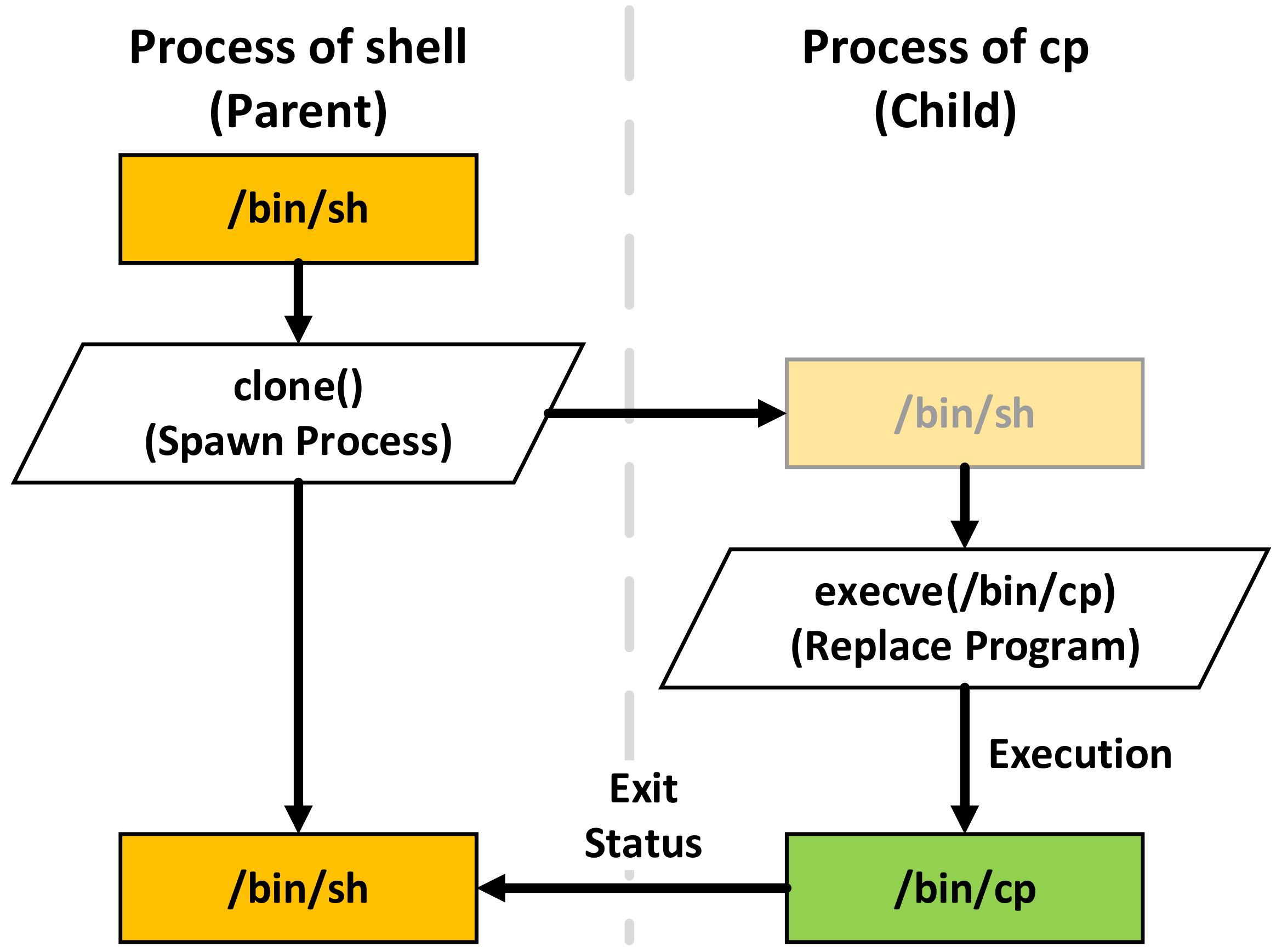}
	
	\vspace{-0.2cm}
	\caption{Spawning of a new process \emph{cp} 
			}
	
	\label{fig:fig-process_execution}
		
\end{figure}

\subsubsection*{\textbf{System Call that Replaces Program's Code}} 
\textit{Execve()} is a  system call that does not have the capability to create a new process, but it often participates in process creation.
When a process calls \emph{execve}, the \emph{execve} executes another program whose path was passed as one of the arguments.
The program of a calling process is thus replaced with a new program code and also stack, heap, and data segments are newly initialized.
Although \emph{execve} can be called by a process at any time, it is usually called after one of the aforementioned system calls; hence after creating a new process, its code is immediately rewritten by the desired code.
An example of process spawning is depicted in Figure~\ref{fig:fig-process_execution}, where Unix utility \emph{cp} is executed. 
First, the process of shell is executed, which then calls a \emph{clone} system call. 
\emph{Clone} creates a copy of the calling process itself (i.e.,~shell).
Next, this copy calls \emph{execve} system call with arguments consisting of \emph{cp} and its parameters.
Once \emph{execve} successfully returns, the code of a new process is replaced by the code of \emph{cp}, and the instruction pointer is set to the first instruction of that code.
When the \emph{cp} process terminates, its return code is passed to the parent process (i.e., shell) to indicate whether \emph{cp} terminated successfully or with an error.

\subsubsection*{\textbf{Flow of System Call Execution}}
\begin{figure}\centering     
    \centering

	\vspace{-0.5cm}    
    \includegraphics[width=0.70\linewidth]{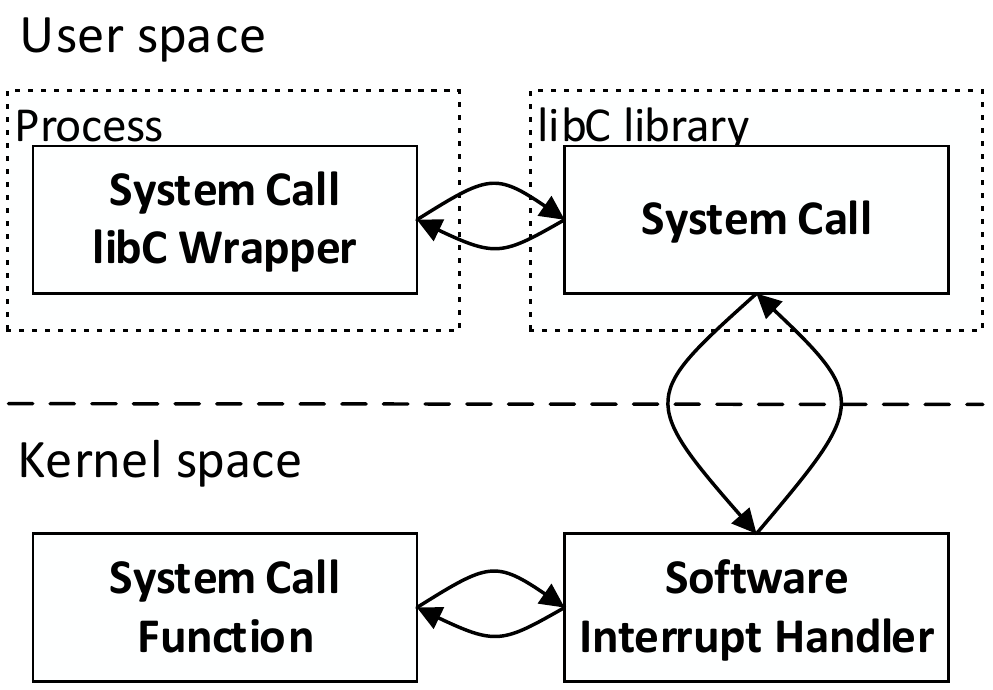}

	\vspace{-0.2cm}
    \caption{Flow of system call execution}
    
    \label{fig:fig-system_call_sequence}
	    
\end{figure}

To create an LKM that intercepts system calls within the kernel space, it is important to understand how system call invocation works.
The flow of a system call execution is depicted in Figure~\ref{fig:fig-system_call_sequence}. When a process invokes a system call, in most cases it does not call the system call directly, but it calls one of the wrappers provided by the libC library.
These wrappers may perform other actions in order to prepare the received data for the actual system call from the Linux kernel.
For example, the libC library provides six different wrappers of \emph{execve} system call.\footnote{\url{https://www.gnu.org/software/libc/manual/html_node/Executing-a-File.html}}
Once the data are prepared, the wrapper calls the system call through a software interrupt that transfers control to the kernel.
Here, the interrupt is caught by the software interrupt handler that may again perform further actions to manipulate the data (e.g., saving all of the processor registers).
Afterwards, the handler looks into the system call table to find the address of the pertinent system call and jumps into that address, initiating the execution of the system call.
Finally, when the system call has been completed, the result is propagated back to the calling process.

\subsection{Loadable Kernel Modules (LKMs)}
An LKM is an object file that can be installed in a Linux kernel in order to extend the kernel's base functionality.
For example, drivers of peripheral devices are implemented as LKMs.
Installation is performed at run time, and once an LKM is installed, its functionality is integrated into the kernel.
The advantages of LKMs are that they can be installed or removed from the kernel at any time, and the Linux kernel does not need to be recompiled to use the LKM.

\subsubsection*{\textbf{Compilation and Installation of an LKM}}
\label{sec:comp_install_lkm}

The compilation of LKMs is different to the compilation of a regular user space program.
Since an LKM is integrated into the Linux kernel upon insertion, the LKM must be compatible with that kernel.
Therefore, an LKM is always compiled against the matching Linux kernel version of the targeted device.
Moreover, it is not only the kernel version that has to match, as the configuration of the kernel features and options must also match (e.g.,~EXT4 file system support, kernel debugging).
Fulfillment of these requirements is simple in conventional machines (e.g., PCs), since a developer can issue the ``\emph{uname~-a}'' command to obtain information about the Linux kernel version as well as the configuration file used for kernel compilation (i.e., "\emph{/boot/}" folder).
However, in the case of IoT devices, the situation is more difficult, since the Linux is usually delivered in a minimalistic and customized form. 
As a result, many necessary files are missing, including the configuration file, therefore the necessary information must be extracted from the device using a different approach (e.g.,~parsing system files \emph{/proc/version} and \emph{/proc/kallsyms}).

Upon insertion of an LKM, a Linux kernel checks the compatibility with the provided LKM using information extracted from the LKM's binary.
However, this check does not cover all of the critical parts that must match.
Although the LKM may pass these checks and be installed in the kernel, there is still no guarantee that it will work properly within the kernel.
If such a situation occurs, the resulting behavior is unpredictable, and in most of the cases, the kernel loses some functionality, freezes, or crashes.
For this reason, it is crucial to match the configuration of the kernel as much as possible.

\section{HADES-IoT}
\label{sec:hades_approach}

We propose a host-based anomaly detection system targeted for IoT devices, called HADES-IoT.
Most of the requirements specified above have been fulfilled since we chose to adopt the LKM approach that utilizes the system call interception technique and, more specifically, intercepts the \emph{execve} system call.
Using the LKM approach, we are able to install HADES-IoT into a Linux kernel at any time; moreover, with this approach there is no need to recompile the kernel.
The only requirement for ensuring that HADES-IoT can run on an IoT device is that HADES-IoT needs to be distributed in binaries that are precompiled (see Section~\ref{sec:hades_compilation}).

\begin{figure}\centering 	
	\centering
		\includegraphics[width=0.70\linewidth]{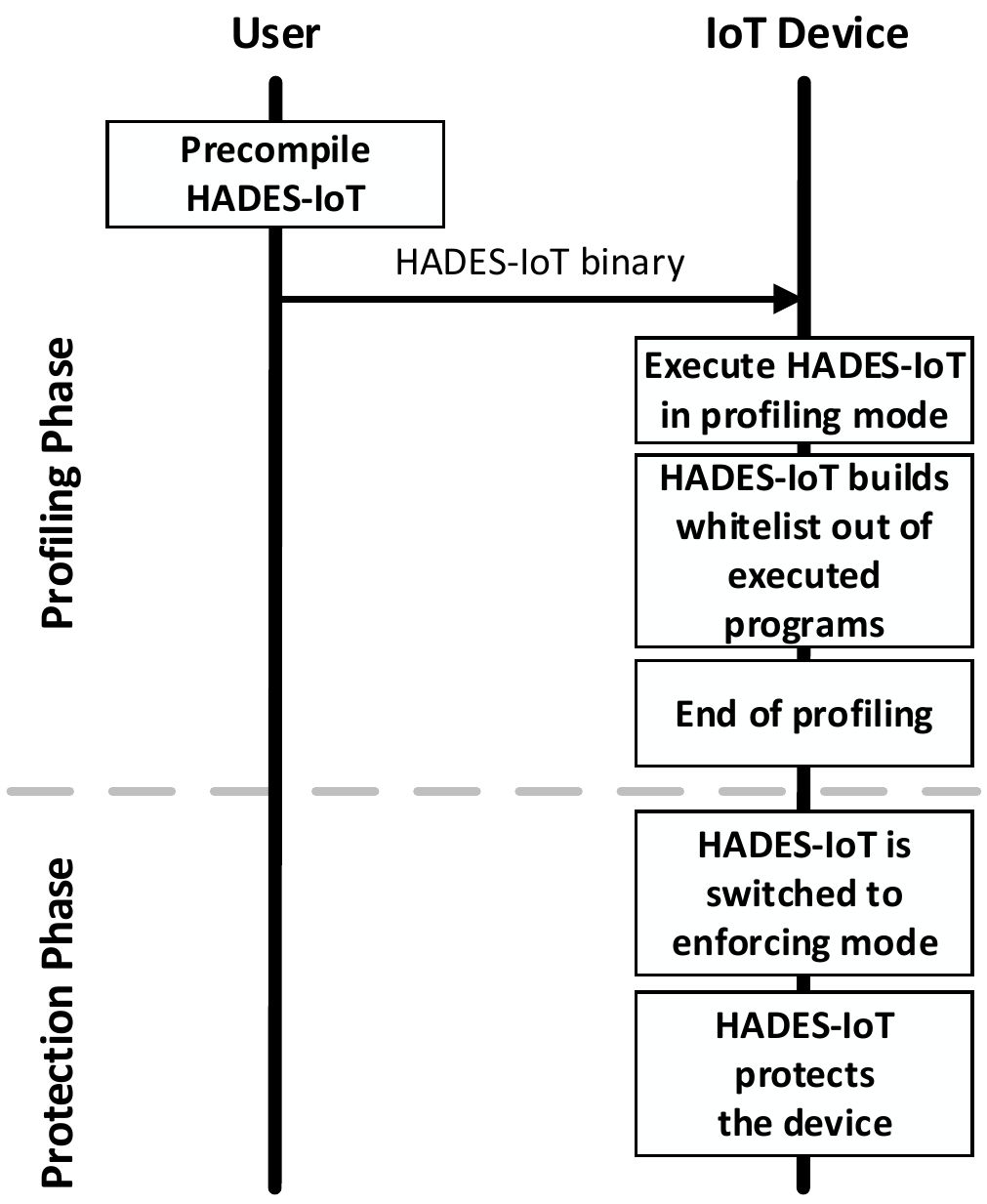}

	\vspace{-0.3cm}
	\caption{Bootstrapping of HADES-IoT: 1) After deployment of HADES-IoT on an IoT device, a profile is extracted and stored in the whitelist, 2) HADES-IoT is switched to the enforcing mode that protects the device using the whitelist.}
		
	\label{fig:fig-hades_bootstrap}
	\vspace{-0.2cm}
	
\end{figure}

HADES-IoT is based on the whitelisting approach. 
The idea of this approach is that only programs that are known to run on an ``uninfected'' off the shelf device are allowed to run.
In order to build a whitelist of benign programs, profiling must be performed once for each device.
This may be viewed as impractical due to the possibility that some benign programs may be missed during profiling.
Nevertheless, HADES-IoT includes a feature that copes with this situation and allows the whitelist to be updated at runtime (see Section~\ref{sec:ext_rem_ctrl_updates}).

\subsection{Bootstrapping}
In the following, we describe the bootstrapping and operation of HADES-IoT, while we distinguish between two modes of our approach: 1) \textbf{profiling mode} and 2) \textbf{enforcing mode}.
HADES-IoT is bootstrapped on a device in two stages (see Figure~\ref{fig:fig-hades_bootstrap}).
First, HADES-IoT is precompiled and delivered to the device, and the kernel's initialization file is modified accordingly to ensure that HADES-IoT is always executed when the device is booted. Once executed, HADES-IoT enters the profiling mode.
In this mode, it monitors and collects information about all calls to \emph{execve}, while the whitelist is updated accordingly.
The profiling stage ends when no new processes are detected during a specified period of time.
We emphasize that during the profiling, also a restart of the device is performed, which enables to update whitelist with all the programs executed at the boot time.
In the last stage of bootstrapping, HADES-IoT is switched to the enforcing mode to protect the device using the whitelist.

\begin{figure}\centering 	
	\centering
		
	\includegraphics[width=0.87\linewidth]{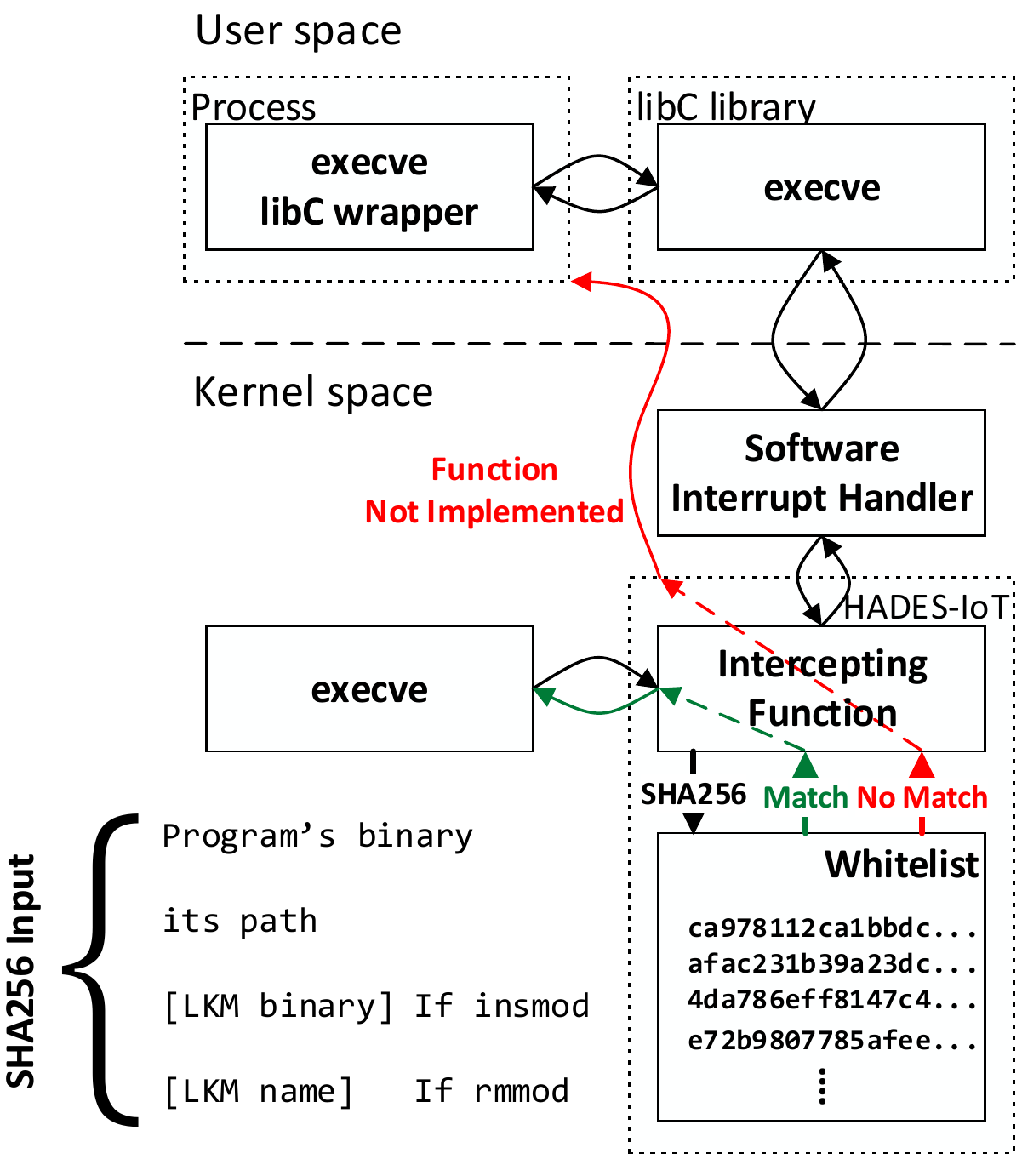}
		\vspace{-0.4cm}
	\caption{Flow of the \emph{execve} system call execution with HADES-IoT installed to the kernel.}
	\label{fig:fig-execve_syscall_seq_hades}
		\vspace{-0.4cm}
	
\end{figure}

\subsection{Detection Process}

The most important feature of the detection process is the intercepting function.
The process of interception is depicted in Figure~\ref{fig:fig-execve_syscall_seq_hades}.
Upon deployment, HADES-IoT locates the system call table and saves the address of the \emph{execve} system call found in the table.
Next, \emph{execve}'s address in the table is replaced by the address of the intercepting function.
This ensures that each time the \emph{execve} is called, the software interrupt handler calls the intercepting function instead of the original \emph{execve} system call.
Once the intercepting function is executed, it first reads the parameters passed to the \emph{execve} system call (i.e., the path of the program to be executed).
Next, the function computes a SHA256 digest out of the program's binary content, its path, and other data, depending on the particular circumstances (see Section~\ref{sec:details_wl_impl}). 
Using the computed digest, the intercepting function looks for a match on the whitelist of all authorized programs.
If a match is found, the process is allowed to run, and therefore the intercepting function performs a call to the original \emph{execve} system call.
However, if a match was not found, the intercepting function returns ``$-ENOSYS$'' error code, which tells the process that the \emph{execve} system call is not implemented.
This naturally terminates the process, thus stopping the execution of any unauthorized (e.g.,~potentially malicious) action.

\subsubsection*{\textbf{Restoration of CPU Context}}
It is important to note that the Linux kernel is not aware of the fact that HADES-IoT changes the address of the \emph{execve} system call in the system call table to the address of the intercepting function.
Therefore, when the \emph{execve} system call is called by a process, the environment is prepared for the execution of this system call.
This means that the intercepting function must act transparently and after performing the authorization check, it must restore any tainted processor register to its original state.
Otherwise, it would lead to an inconsistent state that, in most cases, causes the kernel to freeze or crash.

\subsection{Whitelist Design}
\label{sec:details_wl_impl}
When HADES-IoT is successfully bootstrapped, each call to the \emph{execve} system call is intercepted, followed by a search for a match on the whitelist for a program that is requested to run.
An inefficiently designed search process would cause the IoT device exhibit slow response time, particularly when the device has a large number of periodically spawned processes.
For example, if the whitelist were na\"{i}vely designed as a linked list, the asymptotic time complexity of the search routine would be equal to $O(n)$.
This means that the larger the whitelist is, the longer the imposed delay.
Therefore, to efficiently cope with such a delay, we decided to design the whitelist as a hash table. 
The hash table enables us to reach an asymptotic time complexity of $O(1)$ for a search routine, which means that any delay associated with the search routine remains constant. 

\paragraph{\textbf{IDs in the Whitelist}}
Each item in the whitelist contains ID and represents a program authorized to run on a device.
The ID of an item is a SHA256 digest computed from a program's binary, concatenated with the path to the program, and in certain circumstances, with other additional data.
The reason for such computation is to distinguish symbolic links from the executable they point to.
An example of such an executable that is heavily utilized in Linux-based IoT devices is BusyBox.\footnote{\url{https://busybox.net/}}
BusyBox combines a set of common Unix utilities under a single executable, while particular utilities are accessible through symbolic links. 
Therefore, if we were to compute the digest out of just the binary content of the passed program, we would obtain the digest of BusyBox for all of the utilities.
However, after adding a path element to the digest computation, the resulting digest is different for each of the utilities.

Nevertheless, there are cases in which this approach is not sufficient.
Hence, we have to handle such cases with more fine-grained whitelisting in which additional context dependent data must be added to the input for SHA256 (see Sections~\ref{sec:poss_atks_mal_lkm} and~\ref{sec:poss_atks_hades_uninst}).

\subsection{Tamper-Proof Features of HADES-IoT}
\label{sec:hades_tamper}
In our attacker model, we assume that an attacker is provided with superuser privileges once the IoT device has been compromised.
Therefore, we need to ensure that an attacker who is aware of the presence of HADES-IoT is not able to terminate or modify it.
In the following subsections, we list possible attacks and describe how HADES-IoT protects itself and ensures that it is tamper-proof.

\subsubsection{\textbf{Binary Manipulation}}
\label{sec:hades_bin_manipulation}
When an IoT device is booted, HA\-DES-IoT is deployed to the kernel from its binary file.
Therefore, a possible attack is the deletion of this file.
This would cause the installation of HADES-IoT to fail at boot time.
Similarly, an attacker can move the binary file of HADES-IoT to a different location, again causing the installation of HADES-IoT to fail at boot time.
Finally, instead of removing or moving the binary file, an attacker could attempt to modify HADES-IoT's binary file, which would yield an unpredictable result.
For example, an attacker might modify the kernel version included in HADES-IoT for compatibility checks.
Since the version information contained in HADES-IoT would not match with the version of the kernel running on an IoT device, the kernel would refuse to install HADES-IoT, and thus the device would remain unprotected.
Alternatively, an attacker could corrupt the HADES-IoT binary, which could result in a permanent denial of service due to the crashing or freezing of the kernel.
As mentioned in Section~\ref{sec:comp_install_lkm}, passing the check that takes place during the LKM installation does not guarantee that an LKM is fully compatible with the kernel; hence a faulty LKM can still be installed.

\paragraph{\textbf{Protection}}
To prevent a manipulation of HADES-IoT's binary by an attacker, HADES-IoT loads its binary into the memory on boot and when a restart of a device is requested, HADES-IoT's binary is (re-)written to the storage, regardless of whether the original version was modified or not.
Thanks to this, any malicious modification, removal, or movement of the original binary is prevented.

\subsubsection{\textbf{Loading Malicious LKMs}}
\label{sec:poss_atks_mal_lkm}
A common practice for IoT devices is to install a few LKMs at the boot time.
These LKMs usually represent drivers. However, the utility for installing LKMs (i.e.,~\textit{insmod}) is then included in the whitelist of HADES-IoT. 
This means that an attacker can exploit the utility to install his own LKM in order to ``override'' \emph{execve} interception with a malicious callback function, and thus effectively put HADES-IoT out of the game. 

\paragraph{\textbf{Protection}}
As a protection mechanism, each execution of \textit{isnmod} must be verified against the allowed and known executions, requiring more fine-grained indexing to the whitelist.
In contrast to the whitelist indexing of standard binaries, in the case of \textit{insmod} the index to the whitelist is computed as a cryptographic hash from: 1) binary content of \textit{insmod}, 2) its path, and 3) the binary content of an LKM that is requested to load to a kernel.
This ensures that only known LKMs from the profiling stage are allowed to load again. 
\subsubsection{\textbf{HADES-IoT Uninstallation}}
\label{sec:poss_atks_hades_uninst}

IoT devices often install several LKMs (e.g., drivers) on boot, and therefore they might also uninstall some LKMs during runtime.
When such a case occurs, the utility for uninstalling LKMs (i.e.,~\textit{rmmod}) is included in the whitelist. 
This means that without a prior check, not only the drivers but also HADES-IoT itself can be uninstalled by the attacker.

\paragraph{\textbf{Protection}}
The prevention of this attack is almost the same as in the previous case.
The only difference is that instead of adding the binary content of the LKM to the hash computation, it is sufficient to use the name of the LKM, as the name unambiguously identifies a kernel module that has been already installed.
On the other hand, this solution does not prevent an attacker from uninstalling a kernel module included in the whitelist, and thus disabling some functionality of an IoT device.
We consider this issue out of the scope of this paper, and we plan to address it in our future work.

\subsubsection{\textbf{Init Script Manipulation}}
Another possible attack is a modification of the \emph{init} script.
The \emph{init} script is executed on boot of an IoT device, and it contains commands to configure and prepare the device for use.
HADES-IoT installation is one of the commands included.
Therefore, if an attacker manages to remove the command from the \emph{init} script and subsequently restarts the device, HADES-IoT will not be installed into the kernel on boot.

\paragraph{\textbf{Protection}}
To prevent this attack, as in the case of protecting the modification of HADES-IoT's binary, we propose loading \emph{init} script into the memory of HADES-IoT on boot. 
Therefore, when an IoT device is rebooted, the script is (re-)written to the storage, regardless of whether it was modified or not.

\subsubsection{\textbf{Memory Tampering}}
Memory tampering is another potential attack.
However to prevent this, HADES-IoT takes advantage of the fact that it is integrated into the kernel's memory.
Therefore, user space programs cannot reach the kernel, since accessing the kernel space is forbidden and results in a segmentation fault error.
The only chance for an attacker to tamper with HADES-IoT's memory is to get into the kernel space as well.
Countermeasures that prevent the attacker from loading anything into the kernel were presented in Section~\ref{sec:poss_atks_mal_lkm}.

\subsection{Extensions}

The core functionality of HADES-IoT presented in this paper effectively detects and terminates any unauthorized process spawned on a protected IoT device.
However, in this form, HADES-IoT may be difficult for users to operate.
Therefore, in this section, we present extensions that were incorporated into HADES-IoT in order to make it more convenient to use and to further improve its detection capabilities.

\subsubsection{\textbf{Reporting Subsystem}}
\label{sec:ext_reporting_subsystem}
Although HADES-IoT protects an IoT device and keeps it safe, the owner of the device should be informed about any attempted attack, so he can perform further actions.
Therefore, we introduce a reporting subsystem that informs the owner when an attack has been detected.
The reporting subsystem is optional and contains: 1) an application running on the owner's machine, and 2) a user space application which is deployed along with HADES-IoT on an IoT device.
Once an attack attempt is detected, HADES-IoT immediately notifies the reporting subsystem, which then forwards this information to the owner of the device.

\subsubsection{\textbf{Remote Control \& Whitelist Updates}}
\label{sec:ext_rem_ctrl_updates}
When an IoT device is updated, HADES-IoT must be reprofiled or bootstrapped again (see Section~\ref{sec:dis_firm_updates}).
In the case of reprofiling, HADES-IoT must first be terminated.
However, this termination and reprofiling must be performed in a secure way, not allowing an attacker to take advantage of it.

To address this issue, we further extended the user space application described in Section~\ref{sec:ext_reporting_subsystem}.
This extension allows the application to listen on a specified port, enabling a user to connect through the remote control application to the port.
After connecting, the user must authenticate himself and then the user may issue certain commands to HADES-IoT (e.g., stop, start, profile, protect).
During an update of an IoT device, the user stops HADES-IoT.
Once the update is done, the user instructs HADES-IoT to run in the profiling mode, which causes the whitelist to be rebuilt to reflect the new changes.
After reprofiling, the user sends a command to HADES-IoT in order to enable the protection again.

Since IoT devices rarely provide dependencies for asymmetric cryptography, we propose using Merkle signatures scheme~\cite{merkle1989certified} in order to authenticate messages sent by the owner to the reporting subsystem (see Section~\ref{sec:merkle_tree_sigs}).
Merkle signatures scheme only requires a cryptographically secure hash function, and moreover it provides resilience against quantum computing attacks. 

\subsubsection{\textbf{Signal Monitoring}}
Signals in Linux are software interrupts that inform a process about an event that has occurred.
For each signal, there is a defined default action that a program will perform once the signal is received (e.g.,~stop, continue, or terminate).
A signal can be sent to a process by invoking the \emph{kill} system call.
Another option is to use the \emph{kill} utility.
As sending a signal to a process can lead to its termination, an attacker may take advantage of this by sending the \emph{SIGKILL} signal to a service such as \emph{lighttpd} and cause a denial of service attack.
Nevertheless, when an IoT device is protected by HADES-IoT, the attacker cannot use his own program, since it would be stopped.
Therefore, using the \emph{kill} utility is the attacker's only option.
According to the POSIX standard, the \emph{kill} utility should always be provided as a standalone binary.
If the \emph{kill} utility is never used in the normal profile of an IoT device, HADES-IoT will also detect such an attempt of process termination and prevent it.
Note that the \emph{kill} utility is just a wrapper for the \emph{kill} system call; however, some shells call a built-in function that directly invokes the \emph{kill} system call instead of executing the \emph{kill} utility.
Hence, any malicious termination of a process would not be detected.

To resolve this issue, HADES-IoT also has to intercept the \emph{kill} system call.
This allows HADES-IoT to detect an invocation of the \emph{kill} system call and if it is not authorized, its invocation is prevented.

\section{DEtails of Remote Control}\label{sec:merkle_tree_sigs}
First, we briefly describe Lamport one-time signatures and their aggregation by Merkle signature scheme. 
Then, we explain the integration of this scheme into HADES-IoT for the purpose of authentication of messages sent to the remote control application.
In detail, we describe the deployment of HADES-IoT with this scheme and use case representing secure firmware update.

\subsection{\textbf{Lamport One-Time Signatures}}\label{sec:lamport}
Lamport signature sche\-me~\cite{lamport1979constructing} is a quantum resistant construct of the asymmetric cryptography, which serves for authentication of a single message.
The private key is generated by a cryptographically secure pseudo-random number generator (CSPRNG). 
The private key consists of two pairs of $K$ numbers, each $K$ bits long, where $K$ represents the output size of a cryptographically secure hash function $h(.)$.  
Therefore, the size of the key is $2K^2$ bits (e.g., if $K=256$, then this size is 16kB).
Next, the public key is computed from the private key by making a hash of each number in the key, obtaining the same size and pair-wise structure of the public key as for the private key.

Signature of a message $m$ is created by selecting $K$ numbers from a private key: for each pair of numbers at position $i = \{1, \ldots, K\}$, one number is selected according to the value of a bit at position $i$ of the hash $h(m)$ computed from the message $m$. 
More specifically, the first number of the private key at position $i$ is selected, if the $i$-th bit of $h(m) = 1$, while the second number is selected otherwise.
Hence, a signature contains $K$ numbers and the resulting size of the signature is $K^2$ bits (e.g., if $K=256$, then this size is 8kB).

Verification of the signature associated with $h(m)$ starts by selecting one number from each number pair of the public key, according to the value of each bit in $h(m)$.
Then, the obtained $K$ values are compared to $K$ hash values computed from the signature -- the signature is valid in the case of a match, otherwise, it is invalid.

\subsection{\textbf{Merkle Signature Scheme}}\label{sec:merkle}
Merkle Signatures~\cite{merkle1989certified} extend one-time signature schemes (such as Lamport signatures~\cite{lamport1979constructing}) to support multiple messages.
In detail, $N$ public/private keypairs of one-time signatures are generated by a cryptographically secure pseudo-random function $F(S||i)$, where $S$ represents secret seed, $i = \{1, \ldots, N\}$, $||$ represents string concatenation, and $N$ is equal to the power of 2.
An example of function $F(.)$ is a SHA3 hash function.
Next, the hash value is computed from each public key, and then these $N$ hashes are aggregated by a Merkle tree into a root hash, which represents a master public key associated with all leaf public keys and their corresponding private keys.

Signing of a message $m$ is performed as described in the previous section, but the signature is additionally extended by an authentication path\footnote{Also referred to as \textit{Merkle proof} or \textit{authenticator}.}  associated with a particular leaf in the Merkle tree.
The authentication path consists of $log_2(N)$ hash values and indications of their left/right positions within the Merkle tree.
Given the authentication path and a particular leaf public key, it is possible to verify whether this leaf public key is present at a particular position of the Merkle tree by deriving the root hash value. If the derived root hash value matches the (master) public key, verification is successful.
In sum, the size of the signature is $K^2 + K log_2(N)$ bits (e.g., if $K = 256$ and $N = 2^{15}$, then this size is 8.67kB)

Verification of the signature associated with the message $m$ has two steps:
1) The verification of the Lamport signature is made (as described in Section~\ref{sec:lamport}), 2) if the Lamport signature is correct, then the verification of the expected public/private key pair is made by derivation of the root hash (as described above).

\subsection{\textbf{Integration with HADES-IoT}}
\begin{figure}\centering     
    \centering
        \includegraphics[width=0.90\linewidth]{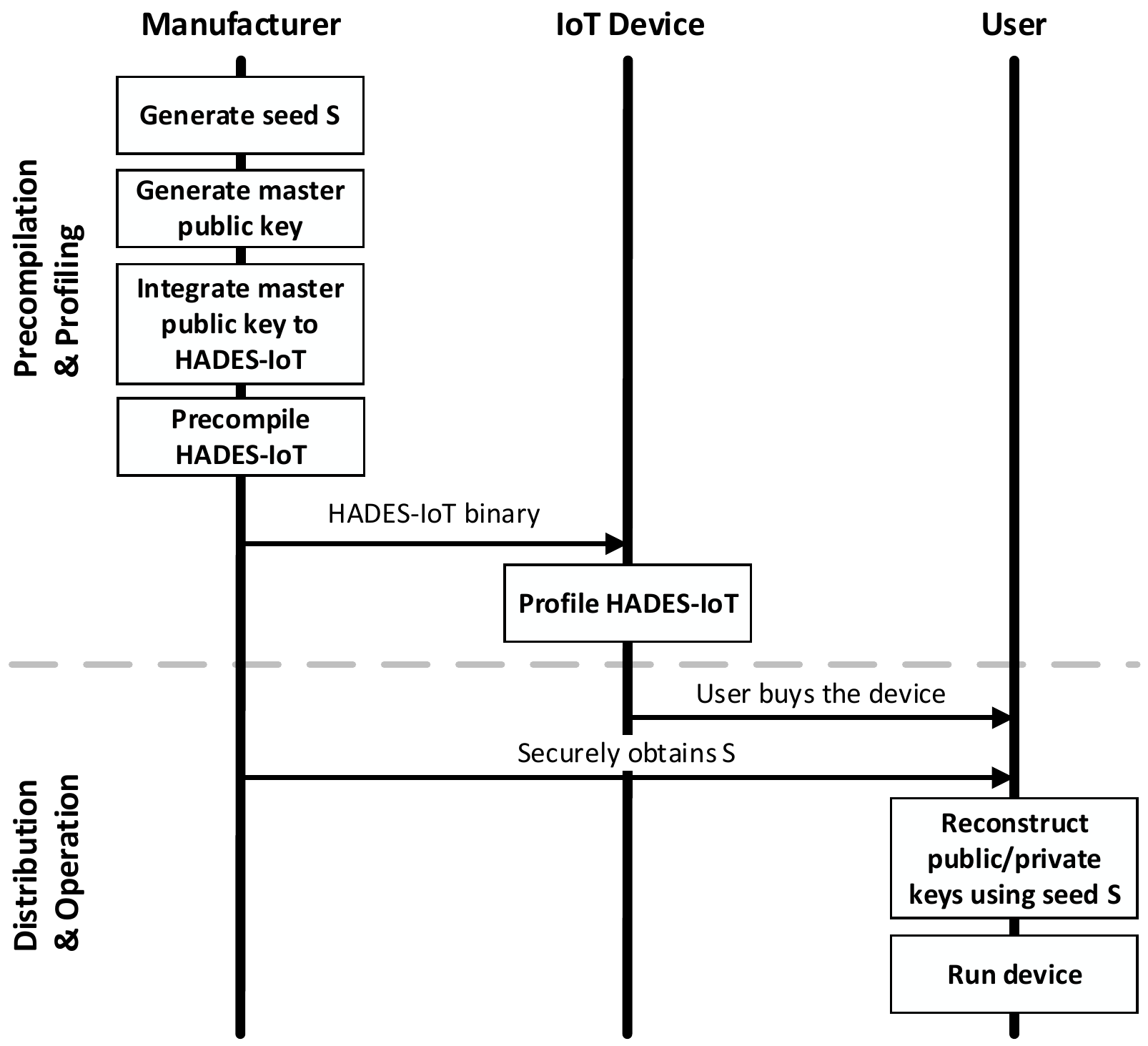}

        \caption{Deployment of HADES-IoT by a manufacturer.}
    
    \label{fig:fig-hades_merkle_man}
        
\end{figure}

\begin{figure}\centering     
            \includegraphics[width=0.55\linewidth]{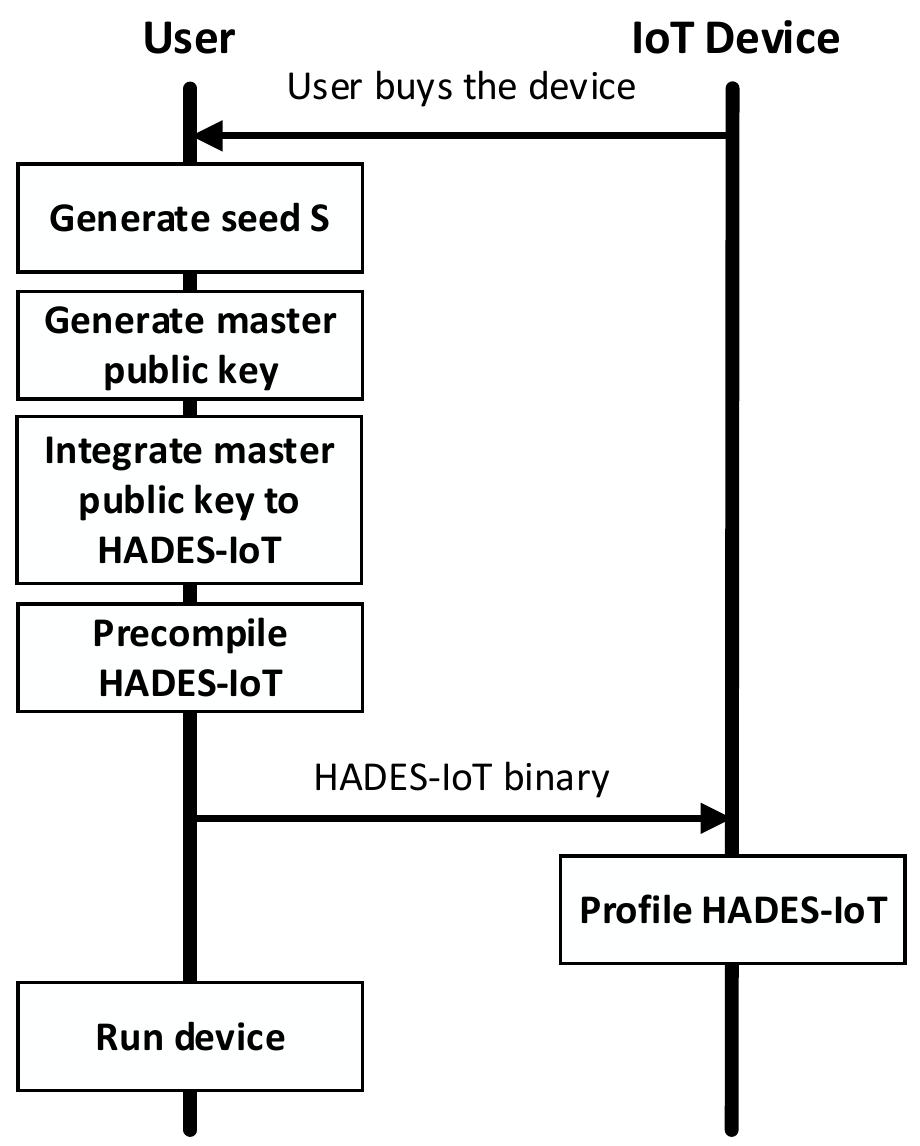}

    \vspace{-0.3cm}
    \caption{Deployment of HADES-IoT by the user.}
        
    \label{fig:fig-hades_merkle_user}
    \vspace{-0.4cm}
    
\end{figure}

We propose to integrate the Merkle signature scheme with HADES-IoT, as it has only minimal requirements on dependencies available on the device -- the only requirement is a cryptographically secure hash function.
In the case when HA\-DES-IoT is deployed by a manufacturer (see Figure~\ref{fig:fig-hades_merkle_man}), the secret seed generation is made by the manufacturer as well.
This may impose issues related to secure delivery of the secret seed $S$ to the user who requires $S$ to generate all public/private keys and reconstruct the Merkle tree.
However, we consider these issues out of the scope of this paper, and we assume that the seed $S$ can be delivered securely.
Note that these issues do not exist when HADES-IoT is deployed by the user (see Figure~\ref{fig:fig-hades_merkle_user}), as the user is the only one who knows the secret seed $S$, and thus all the leaf private keys. 

When all the leaf public keys are generated, the master public key is obtained by computing the root hash out of all the leaf public keys, and then it is integrated into the HADES-IoT's source code. After the integration of the master public key into HADES-IoT and HADES-IoT's deployment on a device, the owner of $S$ (i.e., the user or manufacturer) can send commands (i.e., messages) to the user space application running on the device. 
This application just passes the message to HADES-IoT, which verifies the authenticity of the message, and moreover it verifies whether the ID of the message (i.e., the order of leaf) was not used before -- for this purpose HADES-IoT requires only storing the ID of the last valid message, which avoids reply attacks.
If the verification is successful, HADES-IoT executes the command (e.g., adding a new entry into the whitelist, disabling the protection, etc.).

\begin{figure}\centering     
    \centering  
        \includegraphics[width=0.95\linewidth]{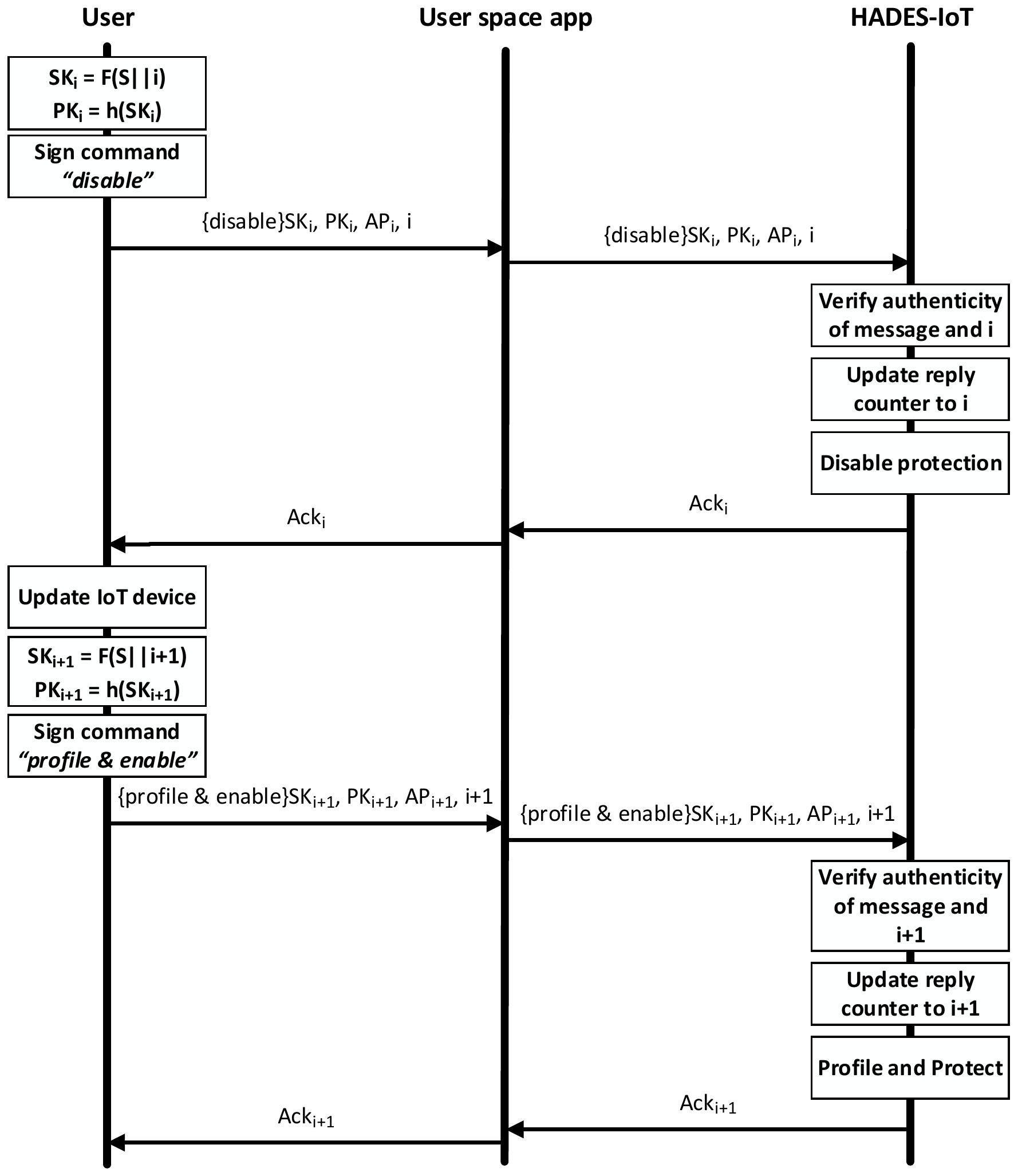}

    \vspace{-0.3cm}
    \caption{Firmware update procedure.}
    
    \label{fig:fig-hades_remote_command}
    
\end{figure}

\vspace{0.1cm}
\textbf{Firmware Update.}
In Figure~\ref{fig:fig-hades_remote_command}, we depict the proposed authentication mechanism in the use case of firmware update performed by the user.
First, leaf secret key $SK_i$ and leaf public key $PK_i$ are computed.
Then, $SK_i$ is used for signing the command that disables enforcing mode of HADES-IoT.
The signed command, together with $PK_i$, authentication path $AP_i$, and ID of the leaf $i$ are sent to the user space application that passes them to the kernel space application of HADES-IoT.
HADES-IoT verifies the authenticity of the message using delivered content and embedded master public key (see Section~\ref{sec:merkle}).
As part of the verification, the ID of the message is compared to the reply counter, and if this ID is smaller or equal, then the authentication fails due to reply attack check, otherwise authentication is successful, and the reply counter is updated.
After authentication of the message, protection by HADES-IoT is disabled, and the user receives an acknowledgment, so he can execute firmware update.
Once the device is updated, the signed command instructing HADES-IoT to perform the profiling and subsequently to enable the protection is sent.
The verification of the authenticity is performed in the same way as in the previous case, and if successful, the protection is enabled.

\setlength{\tabcolsep}{4pt}
\begin{table*}[th]    
    \small{
        \begin{center}
            \catcode`\-=12
            \vspace{-0.5cm}
            \begin{tabular}{l@{\qquad}c@{\quad}c@{\quad}c@{\quad}c@{\quad}c@{\quad}r}
                \hline \noalign{\smallskip}
                \textbf{Device} &
                \textbf{Type} &
                \textbf{\specialcell{Kernel\\Version}} &
                \textbf{\specialcell{CPU\\Architecture}} &
                \textbf{\specialcell{CPU\\Performance\\$[BogoMips]$}} &
                \textbf{\specialcell{Total\\Memory\\$[MB]$}} &
                \textbf{\specialcell{Available\\Memory\\$[MB]$}}\\ \noalign{\smallskip} \hline \noalign{\smallskip}
                Netgear WNR2000v3    & Router    & 2.6.15 & MIPS   & 265.2 &  32 & 16.06 \\ 
                ASUS RT-N16          & Router    & 2.6.21 & MIPSel & 239.2 & 128 & 87.22 \\ 
                ASUS RT-N56U         & Router    & 2.6.21 & MIPSel & 249.3 & 128 & 78.56 \\ 
                Cisco Linksys E4200  & Router    & 2.6.22 & MIPSel & 239.2 &  64 & 18.63 \\ 
                D-Link DCS-942L      & IP Camera & 2.6.28 & ARM    & 534.5 & 128 & 38.75 \\ 
                SimpleHome XCS7-1001 & IP Camera & 3.0.8  & ARM    & 218.7 &  32 & 1.90 \\ 
                Provision PT-737E    & IP Camera & 3.4.35 & ARM    & 218.7 &  32 & 3.88 \\ \hline
            \end{tabular}
            \smallskip
            \caption{IoT devices used in the evaluation.}\label{tab:tab_devices}
            \vspace{-0.5cm}
        \end{center}
    }
\end{table*}
\setlength{\tabcolsep}{1.4pt}

\section{Evaluation}
\label{sec:evaluation}

We implemented a proof-of-concept of HADES-IoT and tested it on seven IoT devices (see Table~\ref{tab:tab_devices}).
In this section, we start by demonstrating that HADES-IoT can be deployed by a manufacturer as well as by an end user.
Furthermore, we experiment with various profiling time periods and determine the minimal amount of time required to extract an accurate profile.
Then, we evaluate the detection performance of HADES-IoT on vulnerabilities exploited by recent IoT malware and measure resource consumption.

\subsection{Precompilation of HADES-IoT}
\label{sec:hades_compilation}

To ensure that configuration options and features of HADES-IoT's LKM match the kernel of a targeted IoT device (see Section~\ref{sec:comp_install_lkm}), we went through a process in which we determined the minimal set of such features and options enabling us to run HADES-IoT.
First, we selected several IoT devices (i.e.,~D-Link DCS-942L, Provision PT-737E, SimpleHome XCS7-1001), cases in which manufacturers provided the kernel source code and configuration.\footnote{Note that in the case of legacy devices, manufactures may not provide it.}
With the selected set of devices, we cross-compiled HADES-IoT against the manufacturer's custom Linux kernels using the default configuration created by the manufacturer. 
We used the manufacturer's toolchain for the compilation of HADES-IoT and verified successful deployment.
In the next step, using Linux kernel archives,\footnote{\url{https://www.kernel.org/}} we downloaded the source code of the generic Linux kernel with the version matching that of the manufacturer. 
Then, we selected the default configuration file of the architecture model closest to the architecture model of the IoT device, modified it to match the manufacturer's configuration as much as possible, and compiled HADES-IoT against the kernel.
After that, we kept selectively removing parts of manufacturer's configuration to determine the minimal options that must match in order to successfully deploy HADES-IoT.
We found that only a few options must match for all of the tested IoT devices, and more importantly, the information about such options can always be extracted directly from the IoT device.

After identifying the minimal configurations, we investigated whether HADES-IoT can be deployed on an IoT device without the support of the manufacturer.
In order to accomplish this, we used a publicly available generic Linux kernel and toolchain for precompilation. We were able to precompile and deploy HADES-IoT with generic resources for all of the tested IoT devices, which indicates that HADES-IoT can be deployed not only by a manufacturer but also by an owner of an IoT device.
Hence, HADES-IoT can protect even legacy devices that are no longer supported by the manufacturers.

\subsubsection{\textbf{Configuration}}
We identified two critical options that must be adjusted in order to run HADES-IoT successfully on all of the tested IoT devices.
First, in Linux configuration the embedded-application binary interface (EABI) must be enabled for successful deployment.
Second, the optimization in compiler must be set to \textit{performance} instead of \textit{size}.
If optimization is set incorrectly, then HADES-IoT will be installed successfully in the kernel but will not function properly.
The details about specific configuration options for the tested devices are presented in the following section.

\subsubsection{\textbf{Device-Specific Configuration Options}}\label{sec:config_options}
Some devices require specific configuration options that must be adjusted in the configuration file for the successful compilation and deployment of HADES-IoT.
In the following, we provide details about such devices, the options they require, and the way how device-specific options can be extracted.

\begin{itemize}
    \item[\textbf{D-Link DCS-942L:}]
    Linux is configured in a preemptive mode on this device.
    Therefore, when configuring Linux options for compilation, the preemptive mode must be enabled.
    Note that in general, the value of this setting can be extracted from file \emph{"/proc/version"}.
            
    \item[\textbf{SimpleHome XCS7-1001:}]
    We found only one specific configuration that must be modified.
    More specifically, the \emph{"CONFIG\_\-ARM\_\-UNWIND"} must be disabled, otherwise, installation of HADES-IoT will fail on compatibility check.
    
    \item[\textbf{Provision PT-737E:}]
    Like the previous case, we must disable the \emph{"CON\-FIG\_\-ARM\_\-UNWIND"} option; however in contrast to the previous device, we may also need to enable the \emph{"CONFIG\_\-FS\_\-POSIX\_\-ACL"} option.
    To determine whether it is necessary to enable this option, the following commands should be issued: \emph{"cat /proc/kallsyms | grep acl"}.
    In cases in which there is a non-empty result, the options must be enabled.
\end{itemize}

\setlength{\tabcolsep}{4pt}
\begin{table*}
	\small{
    \begin{center}
        \catcode`\-=12
		\vspace{-0.4cm}
        \begin{tabular}{l@{\quad}c@{\quad}c@{\quad}c@{\quad}c@{\quad}c}
            \hline \noalign{\smallskip}
            \textbf{IoT Device} & \textbf{\specialcell{Total\\Executables\\Found}} &  \textbf{\specialcell{No. of IDs\\in Whitelist\\1 hour}} &
            \textbf{\specialcell{No. of IDs\\in Whitelist\\2 hours}} &
            \textbf{\specialcell{No. of IDs\\in Whitelist\\4 hours}} &
            \textbf{\specialcell{No. of IDs\\in Whitelist\\1 hour (User Inter).}} \\ \noalign{\smallskip} \hline \noalign{\smallskip}
            \specialcell{Netgear WNR2000v3}    &  526 & 12 & 12 & 12 & 61 \\ 
            \specialcell{ASUS RT-N16}          &  638 &  5 &  5 & 5 & 38 \\ 
            \specialcell{ASUS RT-N56U}         &  375 &  3 &  3 & 3 & 6 \\ 
            \specialcell{Cisco Linksys E4200}  &  399 &  9 &  9 & 9 & 11 \\ 
            \specialcell{D-Link DCS-942L}      & 1256 & 20 & 20 & 20 & 105 \\ 
            \specialcell{SimpleHome XCS7-1001} &  588 &  4 &  4 & 4 & 29 \\ 
            \specialcell{ProVision PT-737E}    &  482 &  5 &  5 & 5 & 9 \\ \hline
        \end{tabular}
		\smallskip
	    \caption{Effect of various profiling periods on the size of whitelist.}    \label{tab:tab_exec_vs_wl}
		\vspace{-0.7cm}
    \end{center}
	}
\end{table*}
\setlength{\tabcolsep}{1.4pt}

\subsection{Profiling Period}
During bootstrapping, HADES-IoT must run in the profiling mode that extracts the profile of an IoT device.
The longer HADES-IoT runs in this mode, the more accurate the extracted profile.
However, it can be inconvenient for a user when the profiling takes too long.
Therefore, we conducted an experiment in which we determined the boundaries on the amount of profiling time needed to obtain an accurate profile.
We experimented with profiling times of one, two, and four hours.
The results of this experiment are presented in Table~\ref{tab:tab_exec_vs_wl}.
We can see that  after one hour of profiling no new processes were found on any of the devices, which means that an accurate profile can be obtained even after one hour of profiling.
On the other hand, there is the possibility that a new program might be executed or a new signal might be sent after a four hour profiling period (e.g., a scheduled job).
However, with the update mechanism described in Section~\ref{sec:ext_rem_ctrl_updates}, any missing program or signal can be added to the whitelist after the profiling phase.
Note that in this experiment, we were not interested in the programs executed during the booting of the device, since these are added into the whitelist regardless of the length of the profiling time.

Next, we measured the difference in the whitelist size when a user interacts with the GUI of a device and cases in which there is no user interaction.
The results show that on some devices, such as \emph{Netgear WNR2000v3} and \emph{D-Link DCS-942L}, the whitelist increases significantly, while on devices, such as \emph{Cisco Linksys E4200} and \emph{ASUS RT-N56U}, the increase is only small.
These results suggest that it is important to interact with the device during the profiling period, otherwise many programs and signals could be missed, leading to a less accurate profile.

Finally, we compared the number of all executables to the number of executables presented in the whitelist.
Table~\ref{tab:tab_exec_vs_wl} shows that each device contains a large number of executables that are never used.
If an attacker compromises the device, none of these executables can be used due to the protection provided by HADES-IoT -- the attacker is strictly limited to the executables in the whitelist.

\subsection{Effectiveness of Detection \& Prevention}
To demonstrate the prevention capabilities of HADES-IoT, we performed several attacks that exploit vulnerabilities used by recent real-world IoT malware.
We describe these attacks in the following. 
\subsubsection{\textbf{Enabled Telnet with Default Credentials \& Mirai}} 
The Mirai IoT malware takes advantage of the fact that many IoT devices connected to the Internet have Telnet open by default, and additionally, that the devices have default credentials configured (SimpleHome IP camera in our set is such a case).
However, with HADES-IoT even such a default misconfiguration does not cause harm, as execution of any unauthorized binary (e.g.,~Mirai binary) is terminated upon its spawning, as witnessed by our evaluation.

\subsubsection{\textbf{[CVE-2017-8225] \& IoTReaper, Persirai}}
CVE-2017-8225 represents a vulnerability of a custom HTTP server that does not properly check access to \emph{.ini} files and allows an attacker to retrieve them; the authentication can be bypassed by providing an empty string for a user name and a password.
By exploiting this vulnerability, the attacker can read the root credentials from the \emph{system.ini} file, as they are stored in plain text.
When the attacker is in possession of the credentials, he can further use it to execute any command on the vulnerable device (e.g.,~Telnet service).
This vulnerability was exploited by Persirai and IoTReaper.

We bootstrapped HADES-IoT on the vulnerable IP camera (i.e., ProVision PT-737E) and executed the exploit.
In the first step of the exploit, the credentials are retrieved by reading the \emph{system.ini} file.
Since this is handled by the HTTP server that is in the whitelist, HADES-IoT allows this action.
In the second step, the remote command is sent through FTP configuration CGI.
According to HADES-IoT's logs, this executes the \emph{chmod} utility, as well as \emph{ftpupload.sh} that executes the command.
However, since none of these executables are in the whitelist, HADES-IoT terminates both of them upon their execution and stops the attack.

\subsubsection{\textbf{[CVE-2013-2678] \& IoTReaper, VPNFilter}}
The CVE-2013-2678 vulnerability enables an attacker to execute an arbitrary command on the affected device.
Therefore, attackers usually exploit this vulnerability to start Telnet service and then deliver the malicious binary on the device.
We used the Metasploit framework to exploit this vulnerability on an unprotected device (Cisco Linksys E4200).
We observed that a \emph{SIGUSR1} signal was sent by \emph{httpd} to \emph{/sbin/preinit} and subsequently, Telnet was enabled.
We then reproduced the attack with HADES-IoT, but the attack failed at the  moment \emph{SIGUSR1} was sent -- the signal did not belong to the whitelist, and thus HADES-IoT stopped it.

\subsubsection{\textbf{[CVE-2014-9583] \& VPNFilter}}
CVE-2014-9583 represents a vulnerability in the \emph{infosrv} service running on ASUS routers.
The service enables an attacker to execute an arbitrary command with root privileges on the vulnerable routers.
Therefore, this vulnerability can be exploited by the VPNFilter IoT malware in order to infect the device and connect it to the botnet.
We tested exploitation of this vulnerability on the ASUS RT-N56U device protected by HADES-IoT in two scenarios: with disabled and enabled Telnet.
With Telnet disabled (default option), an attempt to compromise the device is detected upon exploit execution.
However, in the case in which Telnet is enabled, the attack is detected in its later stage; this means upon malware download or its execution, depending on the user's typical interaction with the device.

\subsubsection{\textbf{TelnetEnable Magic Packet \& VPNFilter}}
Netgear routers allow a user to enable Telnet service via a specifically crafted (magic) packet.
This ``feature'' is exploited by the creators of the VPNFilter malware for its deployment.
We tested the exploitation of this vulnerability on the Netgear WNR2000v3 device.
Like the previous vulnerability, if the Telnet service is disabled, the attack by VPNFilter is detected by HADES-IoT when it begins.
With enabled Telnet, the attack is thwarted as soon as any unauthorized process is spawned (i.e.,~execution of the malware at the latest).

\begin{figure*}[t]
	\centering

	\subfloat[\label{fig:CPU_idle}CPU load (idle)	]{
		\hspace{-0.25cm}
		\includegraphics[width=0.33\textwidth]{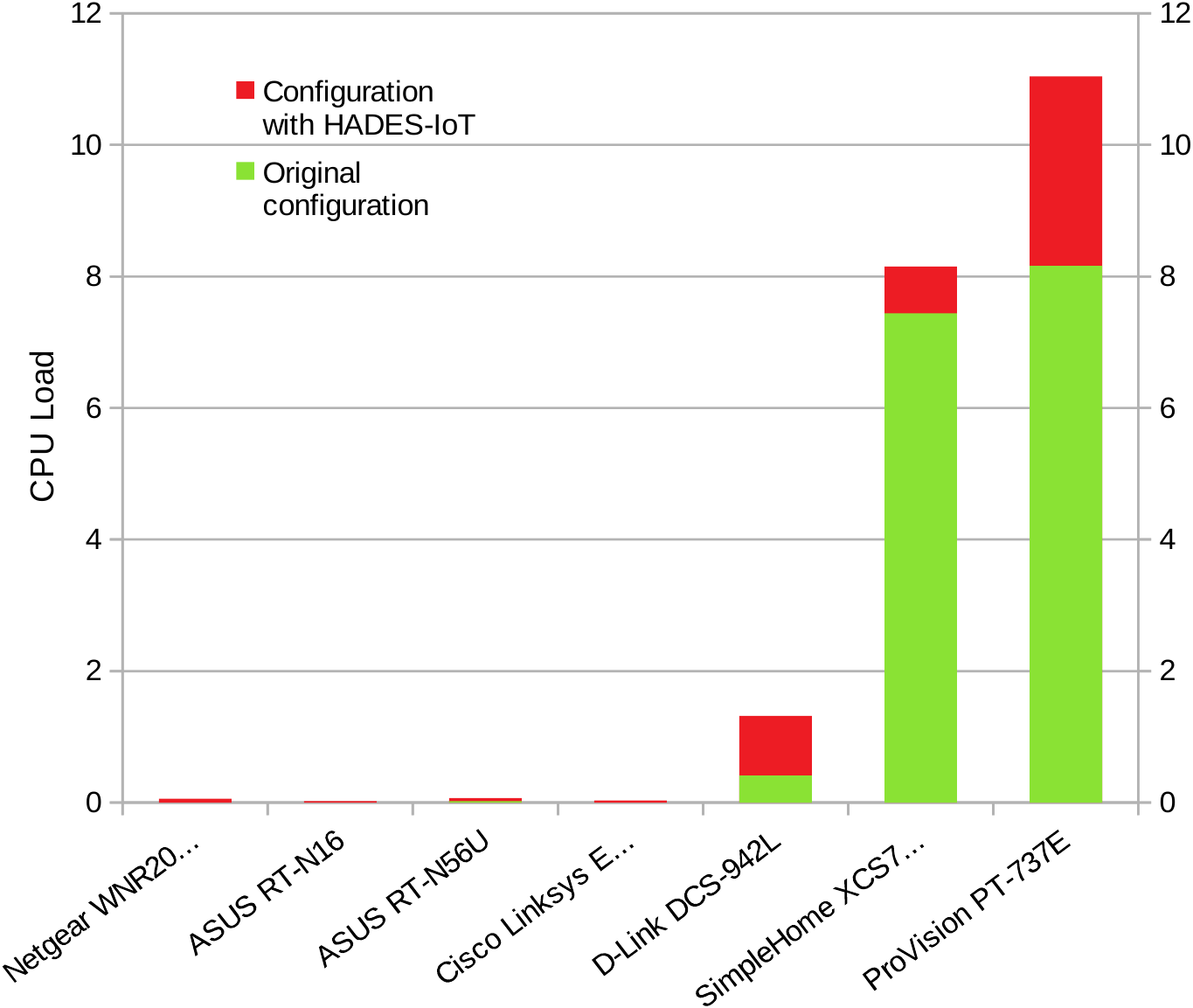} 
	}
	\hspace{0.2cm}
	\subfloat[\label{fig:CPU_busy}CPU load (with user interaction)]{
		\hspace{-0.25cm}
		\includegraphics[width=0.33\textwidth]{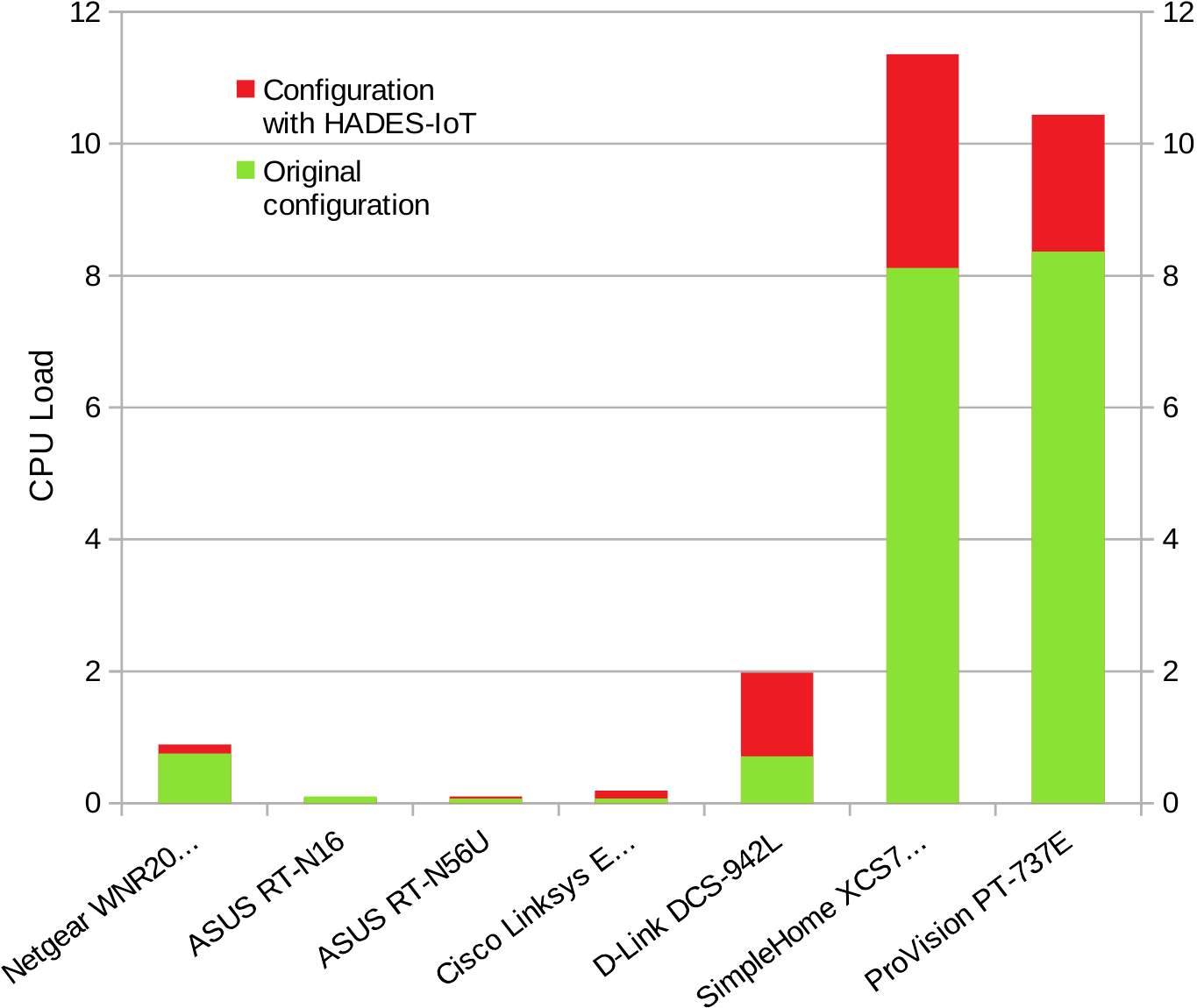} 
	}
	\hspace{0.2cm}
	\subfloat[\label{fig:Memory_overhead}Memory and storage utilization]{
		\hspace{-0.25cm}
		\includegraphics[width=0.31\textwidth]{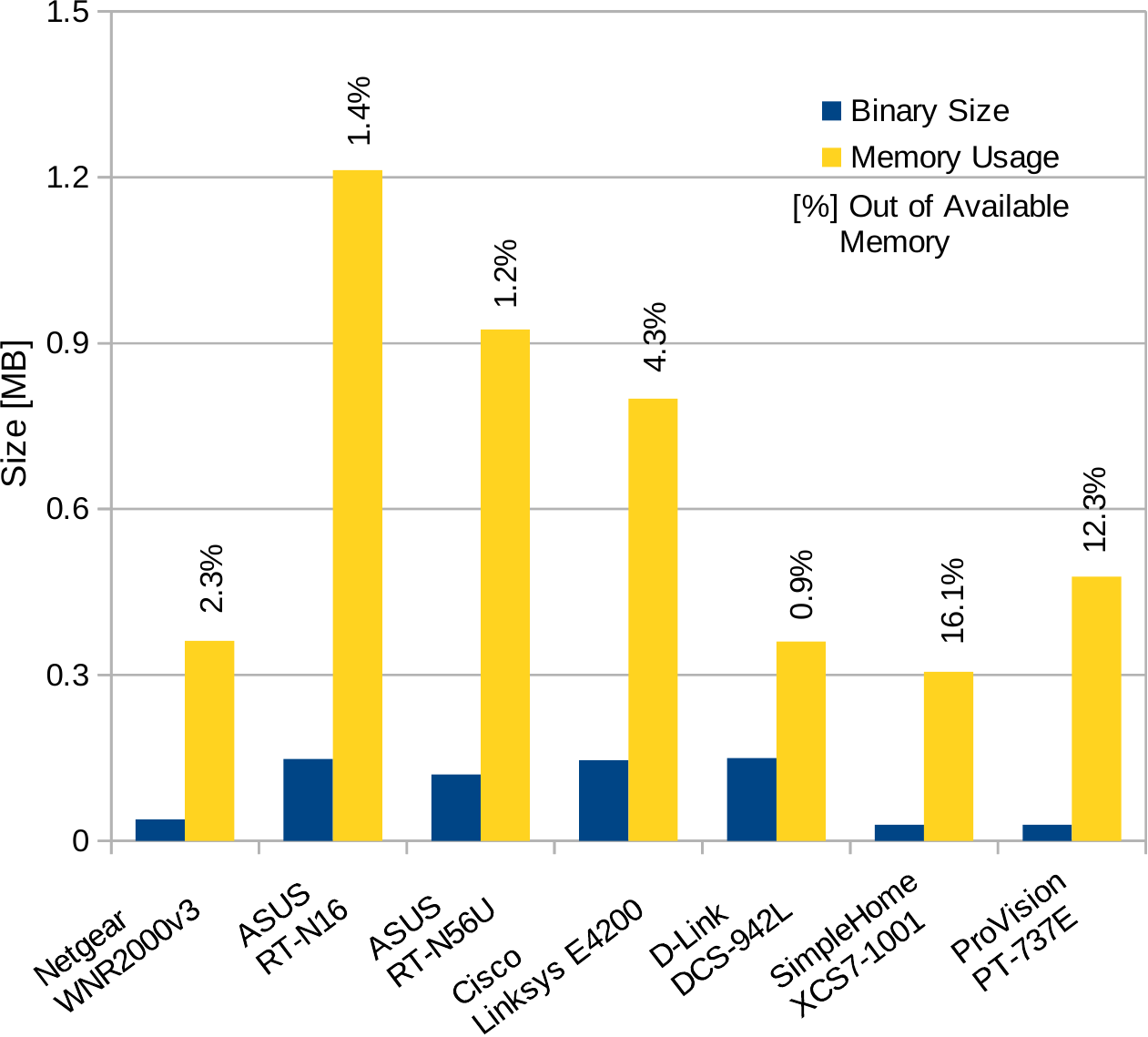} 
	}

	\vspace{-0.2cm}	
	\caption{Utilization of resources by HADES-IoT.}
	\label{fig:overhead}
\end{figure*}

\subsection{CPU and Memory Overhead}

An important aspect of a host-based defense system for IoT devices is low performance overhead.
We conducted an experiment in which we measured the CPU utilization caused by HADES-IoT (see Figures~\ref{fig:CPU_idle} and~\ref{fig:CPU_busy}) and the memory demands (see Figure~\ref{fig:Memory_overhead}).
More specifically, we measured an average CPU load that represents the value of how much work has been done on the system in the previous $N$ minutes~\cite{unix-load} -- in our case we used five minutes.
Although this metric has no clear boundaries, it is the most available option on all IoT devices; common utilities such as \emph{top} do not show LKMs and their resource utilization, hence this information cannot be obtained from them. 
Figure~\ref{fig:CPU_idle} shows that when a device is idle, there is usually only a small amount of overhead.
Note that in some devices (e.g., \emph{ProVision PT-737E}) the CPU load is significantly higher than on other devices, and this is caused by the periodic execution of hundreds processes that, for example, extract various information from the device itself.
Since HADES-IoT checks every executed process, the overhead imposed is higher compared to less active devices -- the imposed overhead comprises $34.6\%$ and $36.9\%$ of the total overhead for the idle case and the case of user interaction, respectively.\footnote{Note that we ignore devices with negligible total overhead, i.e., less than 0.1.} 
Furthermore, we can see in Figure~\ref{fig:CPU_busy} that on some devices (e.g., \emph{SimpleHome} and \emph{D-Link} devices), HADES-IoT imposes greater overhead when a user interacts with the device.
This is caused by spawning additional processes to handle the request, while on other devices the request can be handled by already running processes.

Next, we measured memory and storage demands of HADES-IoT, and the results (presented in Figure~\ref{fig:Memory_overhead}) demonstrate low memory demands in contrast to available space: $5.5\%$ on average.
Regardless of the same source code, the binary size and average memory usage varies for each device.
One of the reasons is that various cross-compilers were used to compile HADES-IoT for each device.
However, what does have the highest impact on the binary size is the configuration of the Linux kernel against which HADES-IoT is compiled.
For example, debug symbols are included in the HADES-IoT's binary due to the kernel configuration, which increases the size of the binary.
Also, note that differences in memory usage across various devices is caused by the introduced tamper-proof mechanisms -- the integrity of HADES-IoT's binary and \emph{init} file that must be loaded into the memory of HADES-IoT.
In detail, these files have a different size on each device, hence loading them into the memory causes different memory consumption for each device.

\section{Discussion}
\label{sec:discussion}

\setlength{\tabcolsep}{4pt}
\begin{table}[b]
	\small{
    \begin{center}
        \catcode`\-=12
		\vspace{-0.45cm}
        \begin{tabular}{ccr}
            \hline \noalign{\smallskip}
            \textbf{Manufacturer} & \textbf{Model} & \textbf{\specialcell{Avg. Time Between\\~Updates $[$Months$]$}} \\ \noalign{\smallskip} \hline \noalign{\smallskip}
            \multirow[t]{5}{*}{ASUS}    & RT-N10E        & 15.03 \\
                                        & RT-N10         &  2.33 \\
                                        & RT-N56U        &  5.28 \\
                                        & RT-N66U        &  2.33 \\
                                        & RT-AC66U       &  2.37 \\ \noalign{\smallskip} \hline \noalign{\smallskip}
            \multirow[t]{5}{*}{D-Link}  & DIR-890L       &  3.21 \\
                                        & DCS-930L       &  7.47 \\
                                        & DAP-1860       &  4.96 \\
                                        & DCS-936L       &  5.77 \\
                                        & DCS-942L       &  6.40 \\ \noalign{\smallskip} \hline \noalign{\smallskip}
            \multirow[t]{5}{*}{TP-Link} & NC220          &  9.10 \\
                                        & TL-WR1043ND V1 &  5.98 \\
                                        & Archer C2300   &  9.17 \\
                                        & Archer D7 V1   &  2.55 \\
                                        & Deco M5 V1     &  1.13 \\ \noalign{\smallskip} \hline \noalign{\smallskip}
            \textbf{Average}            &                &  \textbf{5.54} \\ \hline
        \end{tabular}
		\smallskip
	    \caption{Average time between released updates.}    \label{tab:tab_updates}
		\vspace{-0.6cm}

    \end{center}
	}
\end{table}
\setlength{\tabcolsep}{1.4pt}

\subsection{\textbf{Firmware Updates}}
\label{sec:dis_firm_updates}
To reflect improvements and bug fixes provided by a manufacturer, firmware updates are important to deal with as well.
When an IoT device is updated, the executables on the device are modified, while new ones might be added.
This further means that the behavior in terms of the processes executed may change as well; thus, the profile should be updated in HADES-IoT to reflect these changes.

\vspace{0.1cm}
\noindent \textbf{Na\"{i}ve vs. Advanced Update.}
We consider two types of firmware update procedures on IoT devices.
\textbf{a)} In the na\"{i}ve update procedure, the flash of the device is fully rewritten, which means that HADES-IoT is wiped out.
Therefore, HADES-IoT must be bootstrapped again.
\textbf{b)} In the advanced update procedure, however, only relevant files are changed, while the rest is kept untouched, including the configuration of services.
This makes the update more convenient as there is no need to bootstrap HADES-IoT again, but nevertheless reprofiling must take place.
Note that before updating a device, HADES-IoT must be turned off by remote control (see Section~\ref{sec:ext_rem_ctrl_updates}).

\vspace{0.1cm}
\noindent \textbf{Frequency of Updates.}
Another aspect that must be considered, is the frequency of firmware updates.
If IoT devices were updated very frequently, then it would be inconvenient for an end user to update HADES-IoT after each firmware update. 
To evaluate how often firmware updates are released for IoT devices, we visited the websites of three major IoT manufacturers (i.e., D-Link, TP-Link, and ASUS); then we selected five IoT devices from each manufacturer and inspected the release dates of firmware updates.
In Table~\ref{tab:tab_updates}, we see that the average timespan between update releases is over five months.
Users, however, usually update firmware of their devices infrequently and in some cases they do not update the firmware at all.
Even were users to update their devices each time there is a new firmware update released, we conjecture that updating HADES-IoT does not require a significant amount of extra effort in contrast to the firmware update itself, while taking into account the effective protection provided.

\vspace{0.1cm}
\noindent \textbf{Deployment and Updates by Manufacturer.}
Besides a deployment by the user, HADES-IoT might be deployed by a manufacturer out of the shelf, while updates might be remotely performed by the manufacturer as well.
This would benefit both the manufacturers and end users, as manufacturers could claim that their IoT device is delivered with a support for security solution, 
eliminating time and effort the user would need to spend on updates. Nevertheless, the management and reporting of HADES-IoT could be handled by the user as well, and the user would be able to disable/constraint such a manufacturer's support.

\setlength{\tabcolsep}{4pt}
\begin{table*}[ht]
	\begin{center}
		\catcode`\-=12
		\small{
			\begin{tabular}{l@{\quad}c@{\quad}l@{\quad}c@{\quad}c@{\quad}c@{\quad}c}
				\hline \noalign{\smallskip}
				\textbf{Device} & \textbf{\specialcell{Device\\Type}} & \textbf{Linux} & \textbf{\specialcell{Kernel\\Version}} & \textbf{\specialcell{Unique\\Process}} & \textbf{\specialcell{Non-kernel\\Processes}} & \textbf{\specialcell{Kernel\\Processes\\Included}}\\ \noalign{\smallskip} \hline \noalign{\smallskip}
				-      &  PC & Ubuntu 16.04 (Development) & 4.15.0   & 115 & 118  & 183 \\ 
				- &  PC & Kali 2.0 (Entertainment)   & 4.17.17  & 102 & 132  & 190 \\ 
				Netgear WNR2000v3    & IoT & Custom                     & 2.6.15   &  20 &  30  &  40 \\ 
				ASUS RT-N16          & IoT & Custom                     & 2.6.21   &  30 &  35  &  47 \\ 
				ASUS RT-N56U         & IoT & Custom                     & 2.6.21   &  20 &  31  &  48 \\ 
				Cisco Linksys E4200  & IoT & Custom                     & 2.6.22   &  20 &  20  &  35 \\ 
				D-Link DCS-942L      & IoT & Custom                     & 2.6.28   &  30 &  30  &  62 \\ 
				SimpleHome XCS7-1001 & IoT & Custom                     & 3.0.8    &   9 &  10  &  45 \\ 
				ProVision PT-737E    & IoT & Custom                     & 3.4.35   &   6 &   6  &  43 \\ \hline
			\end{tabular}
		}
	\end{center}
	\smallskip
	\caption{Comparison of IoT devices and PCs. }  \label{tab:tab_iot_vs_pc}
	\vspace{-0.5cm}
\end{table*}
\setlength{\tabcolsep}{1.4pt}

\subsection{\textbf{Deployment on Linux-Based Machines}}
In general, HADES-IoT can run on any Linux-based machine, such as a PC. 
However, it is important to note that HADES-IoT was designed with a focus on IoT devices, having limited resources.
In contrast to IoT devices, PCs with Linux provide us with features, such as \emph{KProbe}, \emph{Auditd}, or \emph{SELinux}, which can be used to devise an approach similar, but easier to develop than HADES-IoT. 
While IoT devices have a rather small and fixed set of applications, the environment of applications on PCs changes more frequently.
To compare differences in these environments, we installed HADES-IoT on two PCs (see Table~\ref{tab:tab_iot_vs_pc}), and we observed a significantly higher number of unique processes compared to IoT devices, while the set of unique processes is not stable.
Another limiting factor for the deployment of HADES-IoT on PCs is the high frequency of updates, which would necessitate frequent reprofiling.
For these reasons, we argue that HADES-IoT is appropriate for IoT devices and is not suitable in volatile environments such as PCs.

\subsection{\textbf{Potential Attacks on HADES-IoT}}
\label{sec:diss_poss_atks_on_hades}

\subsubsection*{\textbf{Buffer Overflow and Code Injection.}}
Buffer overflow attacks take advantage of an unhandled write to a program's buffer and writing the malicious data beyond the boundary of a buffer.
Code injection attacks exploit an improper input validation and enables the injection of malicious code into the program.
In both cases, an attacker can modify a program's execution flow by crafting a specific payload and writing it to the memory.
In general, HADES-IoT does not cover this kind of attack, since it does not check the memory content of a process.
However, it is important to consider the goal of the attacker.
If the attacker performs a code injection attack in order to manipulate the vulnerable service itself, without executing any other program, HADES-IoT does not detect it. 
On the other hand, there are simple protection countermeasures such as address space layout randomization (ASLR) and non-executable stack, which in combination with HADES-IoT can provide full protection.
If the goal of the attacker is to remotely execute a shell command, e.g., with an intention to open a reverse shell or download malicious binaries via the FTP client, the attacker will fail, as these programs would not be included in the whitelist.
Although HADES-IoT is unable to detect the initial phase of such an attack, it will still protect the IoT device in the later phase, when an ``anomalous'' binary is executed.

\vspace{-0.2cm}
\subsubsection*{\textbf{Utilization of Whitelisted Programs}}
Assuming that the attacker knows the set of programs in the whitelist, he can potentially chain these programs to conduct an attack similar to return-oriented programming (ROP)~\cite{roemer2012return}.
Since HADES-IoT only checks the parameters of programs in specific cases, such an attack would not be detected. Note that this attack is possible given HADES-IoT's design, hence it cannot be fully prevented;
however, it can be mitigated by more fine-grained logic of the whitelist.

\subsection{Possible Extensions}
\label{sec:diss_extensions}
In this section, we present possible extensions aimed at improving practical features of HADES-IoT.
These features are not part of the proof-of-concept, as they do not improve the detection performance.

\subsubsection*{\textbf{Malware Collection}}
In order to hide traces, it is a common practice for malware to delete itself after its execution.
Since HADES-IoT pauses program execution in its initial stage, we are able to retrieve the binary of malware before it is (potentially) deleted.
Therefore, in addition to the reporting subsystem (see Section~\ref{sec:ext_reporting_subsystem}), HADES-IoT can be enhanced by the capability of collecting and reporting suspicious binaries.
 Such a capability would enable us to collect the latest malware samples. 

\subsubsection*{\textbf{Automated Extraction of Configuration}}

To ensure the compatibility of HADES-IoT with a Linux kernel on a particular IoT device, some configuration options must be extracted from the device (see Section~\ref{sec:comp_install_lkm}) and stored in the kernel configuration file.
Since the configuration file consists of information about all of the options set, it can be parsed and adjusted in an automated fashion instead of being adjusted by manual configuration.
Therefore, the kernel configuration process performed at the user's computer could be automated by running a script that would connect to an IoT device, issue necessary commands for extraction of the important data, and based on that, modify the stock configuration file of a vanilla Linux.
Next, the configuration file obtained would be used for the compilation of the required kernel's parts, enabling proper compilation of HADES-IoT.

\section{Related Work}
\label{sec:related_work}

In this section, we discuss works aimed at host-based intrusion detection systems for IoT devices. 
An anomaly-based approach is provided by Yoon et al.~\cite{yoon2017learning}; the authors present a lightweight method based on the distribution of system call frequencies.
By utilizing a cluster analysis, they first learn the benign execution context, and then a device is monitored in real-time to detect anomalies.
However, the authors only consider attacks that alter system calls in benign programs and use only one sample for their evaluation (i.e., \texttt{Motion}).
A lightweight IDS that focuses on smart meters is proposed in~\cite{tabrizi2014model}; this research is based on system call sequences, where the benign program is represented by a finite state machine (FSM).
The system calls are captured by the \emph{strace} utility and stored in a log file.
Using \emph{strace} for system call collection, the authors incur only $1$\% of performance overhead.
On the other hand, the second component compares the captured system calls stored in the log with the FSM, hence it is more resource intensive, and the authors execute it only every $10$ seconds.
To lower the resource demands, the authors consider only those system calls that an attacker can possibly utilize during the attack.
Using this approach, the IDS incurs overhead of only $4$\%.
However, the presented IDS is designed to be used on smart meters only, so it is trained on a single executable.
On top of that, it requires the addition of annotations to the smart meter software by its developers.
The work of Agarwal et al.~\cite{agarwal2017detecting} presents a concept for anomaly detection that uses context-sensitive features based on Ball-Larus path profiling.
However, this approach requires the source code be instrumented, which facilitates the recording of function calls during execution.
This study includes just a preliminary evaluation of the overhead inflicted to two programs -- \textit{tcpecho} and \textit{consumer health IoT gateway}, and the detection performance is not evaluated.
An et al.~\cite{an2017behavioral} propose behavioral anomaly detection aimed at home routers.
This research employs three semi-supervised algorithms (i.e.,~principal component analysis, one-class SVM, and a na\"{i}ve detector based on unseen n-grams) utilizing captured kernel-level system calls to determine whether a device has been compromised.
The system calls used for training are extracted from a device using the \emph{ftrace} utility.
In the experiments, the trained classifiers are evaluated on two IoT malware families -- one variant of MrBlack and four variants of Mirai.
The results show that all three algorithms employed achieved a 100\% detection rate with a low false alarm rate, but the overhead inflicted by the approach is not evaluated.
The downside of the approach is that full kernel recompilation with enabled \emph{ftrace} support is required.
Su et al.~\cite{su2018lightweight} present a lightweight image recognition technique for malware classification.
The proposed approach transforms a program's binary into images of 64x64 pixels.
Such images then serve as input to a convolutional neural network that determines whether the analyzed program is malicious or benign.
To evaluate the performance, the authors used malware captured by the IoTPOT honeypot~\cite{pa2016iotpot}.
They achieved 94\% accuracy for two-class classification and 81\% accuracy for three-class classification (i.e.,~benign, Mirai, or Gafgyt).
However, the authors admit that their approach is susceptible to complex code obfuscation.
From the industry sector, NEC announced that it has developed tamper detection technology for IoT devices~\cite{nec2018tamper} that leverages ARM's security technology, \emph{TrustZone}.
The proposed approach should be able to detect tampering with a device in just six milliseconds.
The detection method is based on memory inspection, and the vendor claims that their approach checks only 2kB memory.
The article further states that the proposed detection method is able to discover tampering with a device not only during the operation of the device, but on its first activation as well (e.g., detection of supply chain attacks).
Note that thorough explanation of the approach is not provided and evaluation is also lacking. 

\section{Conclusion}
\label{sec:conclusion}
In this paper, we proposed HADES-IoT, a host-based anomaly detection system for IoT devices, which provides proactive detection and tamper-proof resistance. 
HADES-IoT is based on whitelisting and utilizes system call interception which is performed within the loadable kernel module (LKM).
Thanks to the LKM, HADES-IoT gains complete control of the execution of all user space programs, and any execution of an unauthorized binary can be thwarted when it starts.
HADES-IoT is lightweight in terms of its size, memory, and CPU demands.
Computational overhead is only influenced by the number of spawned processes on the device, but not by operations with the whitelist -- 
searching in the whitelist has a constant time complexity. In the evaluation, we showed that extraction of an accurate device profile can be performed in an hour; HADES-IoT also demonstrated 100\% effectiveness in the detection of five kinds of attacks.


\bibliographystyle{ACM-Reference-Format}
\bibliography{hades} 

\end{document}